\documentclass[aps,prb,twocolumn,amsmath,amssymb,showpacs,superscriptaddress,notitlepage,longbibliography]{revtex4-1}
\usepackage{subfigure}
\usepackage{float}
\usepackage{color}
\usepackage{graphicx}
\usepackage{dcolumn}
\usepackage{bm}

\usepackage{amssymb}
\usepackage{esint}
\usepackage{mathrsfs}
\usepackage{booktabs}
\usepackage{multirow}
\usepackage{makecell}
\usepackage{verbatim}

\usepackage[colorlinks,linkcolor=red,anchorcolor=black,citecolor=blue]{hyperref}

\begin{document}
\title{Protocol for detecting the nonlocality of the multi-Majorana Systems}

\author{Bai-Ting Liu}
\thanks{These authors contributed equally to this work.}
\affiliation{State Key Laboratory of Low Dimensional Quantum Physics, Department of Physics, Tsinghua University, Beijing, 100084, China}
\affiliation{Frontier Science Center for Quantum Information, Beijing 100184, China}

\author{Peng Qian}
\thanks{These authors contributed equally to this work.}
\affiliation{Beijing Academy of Quantum Information Sciences, Beijing 100193, China}

\author{Zhan Cao}
\affiliation{Beijing Academy of Quantum Information Sciences, Beijing 100193, China}

\author{Dong E. Liu}
\email{Corresponding to: dongeliu@mail.tsinghua.edu.cn}
\affiliation{State Key Laboratory of Low Dimensional Quantum Physics, Department of Physics, Tsinghua University, Beijing, 100084, China}
\affiliation{Frontier Science Center for Quantum Information, Beijing 100184, China}
\affiliation{Beijing Academy of Quantum Information Sciences, Beijing 100193, China}
\affiliation{Hefei National Laboratory, Hefei 230088, China}

\begin{abstract}
    Majorana zero modes (MZMs) are non-Abelian quasiparticles with the potential to serve as topological qubits for fault-tolerant quantum computing due to their ability to encode quantum information nonlocally. In multi-Majorana systems configured into two separated subsystems, nontrivial quantum correlations persist, but the presence of trivial Andreev bound states (ABSs) can obscure this nonlocality if MZM preparation fails. To address this, we propose a protocol using an entanglement witness based solely on parity measurements to distinguish the nonlocal characteristics of MZM systems. Our framework, which is experimentally implementable, achieves a detection probability of approximately 18\% in a 6-site system and demonstrates robustness under environmental noise, albeit with a reduced detection rate in the presence of quasiparticle contamination.

\end{abstract}

\maketitle

\section{Introduction}\label{sec:intro}
Majorana zero modes (MZMs) are self-conjugate quasiparticles with non-Abelian statistics, drawing considerable attention for their potential in topological quantum computing (TQC)~\cite{ReadGreen,1DwiresKitaev,toricKitaev,NayakRMP}. 
Unlike bosons or fermions, exchanging MZMs transforms the wave function according to topological properties. 
Kitaev’s TQC proposal~\cite{toricKitaev} harnesses the robustness of topological states, offering stronger protection against local noise than conventional qubits. 
A measurement-only TQC approach~\cite{MeasurementOnlyTQC} (with a detailed nanowire-based design in~\cite{Karzig2017Scalabledesigns}) streamlines the process by eliminating the need to move MZMs, focusing instead on joint-parity measurements.

The search for Majorana zero modes (MZMs) has attracted intense interest in condensed matter physics due to their unconventional exchange statistics, governed by topological properties rather than standard bosonic or fermionic rules. Theoretically, MZMs can emerge in topological superconductors, such as one-dimensional (1D) wires or vortex cores in two-dimensional (2D) superconductors, where they localize at the boundaries of topological defects~\cite{ivanov_non-abelian_2001,fu_superconducting_2008,LutchynPRL10,Oreg_helical_2010,DasSarma2010,Qi_Zhang2011}.

Experimentally, 1D superconductor–semiconductor heterostructures have been central to MZM research. In 2012, Kouwenhoven’s group at Delft University reported MZMs in InSb nanowires proximitized by a conventional superconductor~\cite{mourik_signatures_nodate}, observed as a zero-bias conductance peak under an external magnetic field. Subsequent experiments confirmed similar zero-bias peaks at topological semiconductor–superconductor interfaces, especially in materials with strong spin–orbit coupling (e.g., InSb/InAs–Al or PbTe–Pb)~\cite{Rokhinson2012,nichele_scaling_2017,suominen_zero-energy_2017,chen_experimental_2017,albrecht_exponential_2016,finck_anomalous_2013,das_zero-bias_2012}.

Beyond 1D nanowires, 2D systems, particularly fractional quantum Hall (FQH) states, may host MZMs at vortex cores. In $\nu=5/2$~\cite{sarma_majorana_2015} and $\nu=2/3$~\cite{park_manifestation_2024} FQH systems, the edge modes can support non-Abelian anyons.

Another promising platform that can host MZMs is the iron-based superconductors. The vortex in iron-based SC could bound zero-energy modes. Around 2018, experiments in Fe-based superconductors revealed zero-bias bound states~\cite{yin_observation_2015}, subsequently identified as vortex Majoranas~\cite{zhang_observation_2018,PhysRevX.8.041056,wang_evidence_2018,yuan_evidence_2019,machida_zero-energy_2019,zhu_nearly_2020}. Recent work shows that these vortex Majoranas can form lattices on Fe-based superconductor surfaces~\cite{chiu_scalable_2020,li_ordered_2022} and can be braided~\cite{posske_vortex_2020,stenger_simulating_2021}.

Before realizing topological quantum computation through MZMs, substantial efforts are required to identify and confirm their key properties, particularly their nonlocal nature. In superconductor--semiconductor hybrid systems, disorder often leads to the emergence of local Andreev bound states (ABSs). Although ABSs exhibit many local characteristics similar to MZMs---including the zero-bias peak (ZBP)---they lack the nonlocal topological properties essential for quantum information applications.
Local experimental techniques face significant challenges in differentiating ABSs from genuine MZMs. Nevertheless, our study provides new insights from a quantum information perspective, potentially helping to distinguish these two types of bound states by examining their nonlocal properties.

The entanglement witness is one way to detect quantum entanglement~\cite{Horodecki09}.
By measuring the expectation values of a witness operator that is usually local or easy to measure, one can detect the presence of entanglement in both bipartite and multipartite quantum system~\cite{Sperling13}. 
If the resulting value is less than or equal to a certain threshold, then the system is entangled. Otherwise, the system is deemed separable.
Such a threshold is very similar to the famous Bell's inequality. Actually Bell's inequality (and the CHSH inequality) is a particular non-optimal witness of entanglement~\cite{Toth05,witnessBell2005}.
The entanglement witness can also be related to other quantity for measuring entanglement, such as negativity, entanglement of formation, and the robustness of entanglement~\cite{Eisert_2007}.
Moreover, the method of witnessing can also be used to detect other quantities that measure the quantum resource, for example, quantum coherence~\cite{coherenceWitness2021} and negativity of Wigner functions~\cite{chabaud_witnessing_2021}.

In~\cite{romito2017ubiquitous}, the authors demonstrate that MZMs can violate Bell or CHSH inequalities as a consequence of their quantum nonlocality.
Taking a bipartite system composed of six MZMs (in which any allowed pairing of MZMs is nonlocal) as an illustrative example, the authors propose that weak measurement protocols can be employed to detect the resulting nonlocality that gives rise to CHSH inequality violations.

In our paper, we use a method similar to the entanglement witness to witness the MZM paring. The idea is to design a witness operator for the quantum correlation in MZM pairs, which distinguishes it from topologically trivial states.
Our method can carry out the task with several parity measurements, focusing only on the quantum information aspect.

The paper is structured as follows. 
In Section~\ref{sec:pre}, we provide a brief overview of the physical systems discussed in the article.
The nonlocality exhibited by a pair of MZMs when they appear in pairs is a crucial feature in our study.
This characteristic enables us to investigate the nonlocal nature of MZM systems and and distinguish between MZM pairs and unwanted localized states, specifically ABS.
In Section~\ref{sec:tunn}, the physics of tunneling between quasiparticles on hybrid nanowires and the quantum dots (QDs) is also introduced.
The tunneling physics of QD and the quasiparticles is an important tool we use to measure the witness for MZM.
And in Section~\ref{sec:witness}, we demonstrate in detail the witness we designed aiming at detecting the nonlocality of MZMs.
The different situations of the MZM-paired state and ABS are also discussed here.
In Section~\ref{sec:protocol}, we propose a protocol to distinguish MZMs from Andreev bound states (ABS). 
It utilizes an operator similar to an entanglement witness, which is commonly employed to differentiate between entangled states and separable direct product states.
The protocol is presented step by step on a simple toy model, a system of six interconnected topological superconducting nanowires.
And in Section~\ref{sec:calculation}we also analytically computed the performance of the selected witness operator in this simple model.
Then in Section~\ref{sec:MZMerror} we discuss the performance of the witness operator when MZMs are affected by noise. Specifically, we focus on the noise form known as quasiparticle poisoning, which cannot change the locality of MZMs.
The quasi-particle poisoning only changes the parity of MZM pairs.
Some preliminary numerical results for the performance of witnessing operator under quasi-particle noise are presented in this section. 
Section~\ref{sec:conclusion} gives a brief summary of the article and some future perspectives on open questions.

\section{\label{sec:pre}Preliminaries}
\subsection{\label{sec:nonlocal} Majorana nonlocal encoding}
To differentiate MZMs from the non-desirable regular fermionic state like Andreev bound states (ABSs), it is essential to focus on their fundamental disparity. In this section, we elucidate how the pairing of MZMs encodes information in a nonlocal manner, contrasting sharply with ABSs. This critical distinction forms the basis of our proposed protocol.

MZMs represent a unique subset of non-Abelian anyons, manifesting as zero-energy bound states within specific topological superconductors. From an operatorial perspective, each MZM is characterized by a Hermitian operator $\gamma_i$, which adheres to the algebraic relation:
\begin{equation}
    \left\{\gamma_{i}, \gamma_{j}\right\}=2 \delta_{i j}, \gamma_{i}=\gamma_{i}^{\dagger}
\end{equation}
where $i$ and $j$ denote spatial locations conducive to the existence of Majorana zero modes.

The operators associated with MZMs at a pair of spatial sites ${i,j}$ can be expressed through the linear combination of fermionic creation and annihilation operators at these sites, as follows:
\begin{equation}\label{eq:MZM_op}
\begin{aligned}
    c_{i j}^{\dagger}=(\gamma_{i}+\mathrm{i} \gamma_{j}) / 2&, c_{i j}=(\gamma_{i}-\mathrm{i} \gamma_{j}) / 2\\
    \gamma_{i}=(c_{i j}^{\dagger}+c_{i j})&, \gamma_{j}=-\mathrm{i}(c_{i j}^{\dagger}-c_{i j})
\end{aligned}
\end{equation}
Moreover, the parity operator spanning the pair of sites ${i,j}$ for MZM configurations is delineated as:
\begin{equation}\label{eq:Parity}
P_{ij} = \mathrm{i}\gamma_i \gamma_j = 1- 2 c_{ij}^\dagger c_{ij}
\end{equation}
Within this framework, two distinct parity states emerge at a given pair of sites, namely the odd-parity states $|0_{ij}\rangle$ and the even-parity states $|1_{ij}\rangle$. It is  important to note that each parity category encompasses two degenerate eigenstates. These states lay the foundation for representing the computational basis states $|0\rangle$ and $|1\rangle$ in the realm of quantum computation.

In the operational realm of topological quantum computation, leveraging MZMs with conserved parity is essential. Thus, to construct a topological qubit, it necessitates the inclusion of at least two pairs of MZMs, each distinguished by differing parities, as elucidated in~\cite{Beenakker2020}. For configurations that extend beyond two sites, the method of MZM pairing significantly influences the selection of the computational base set.

\begin{figure}[htbp]
    \centering
    \includegraphics[width=0.8\linewidth]{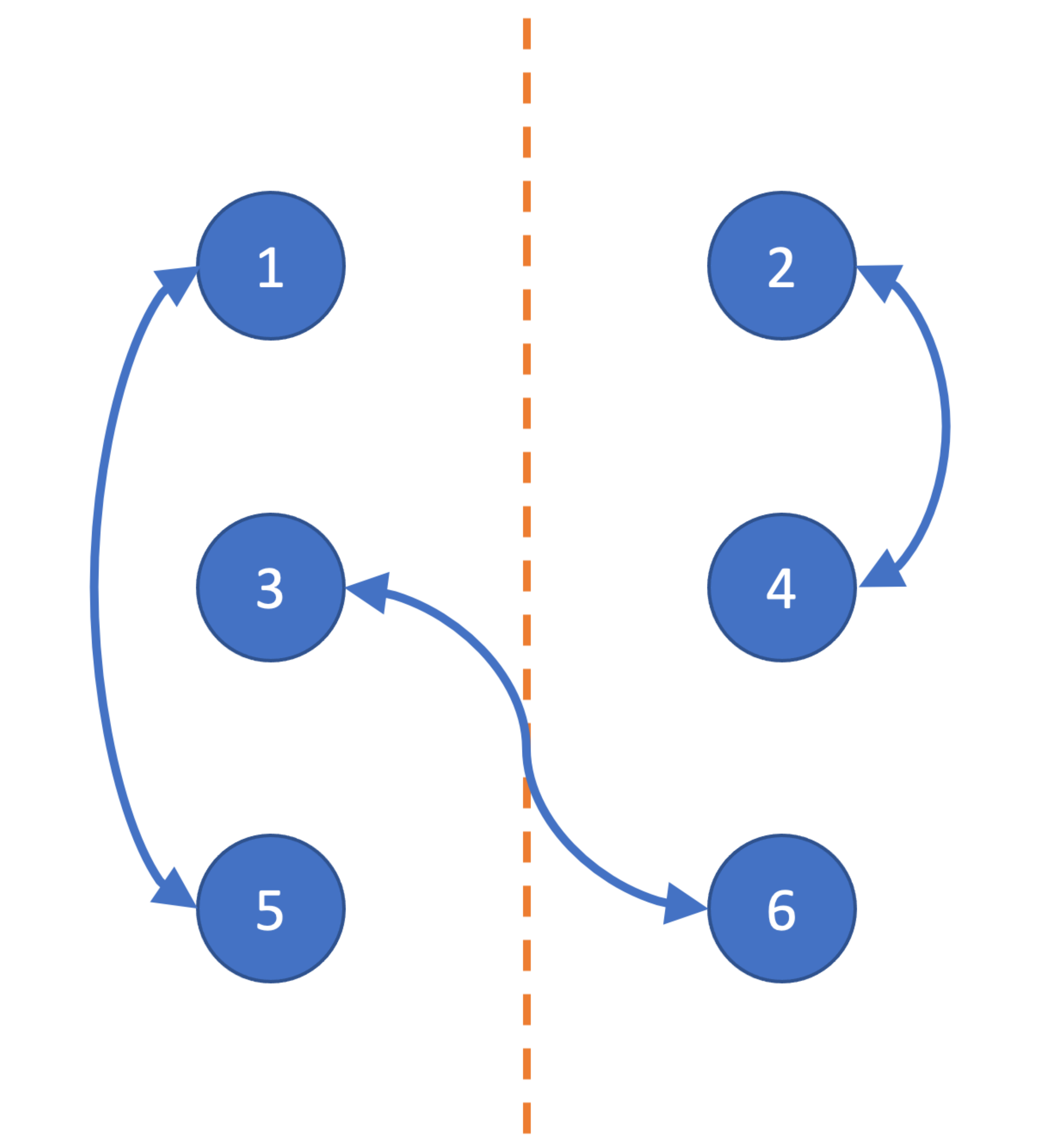}
    \caption{\label{fig:sites} The 6 sites are labeled as $1,2,\dots,6$ and partitioned into two parts: $1, 2, 3$ and $4, 5, 6$, as shown by the grey dashed line.
    On these 6 sites there are 3 MZM pairs, as shown by the light pink dashed lines.
    Among these pairs of MZM pairs, there would always be at least one pair that spans the two subsystems, no matter how we pair up the sites.}
\end{figure}

\begin{figure*}[tb]
    \centering
    \includegraphics[width=\textwidth]{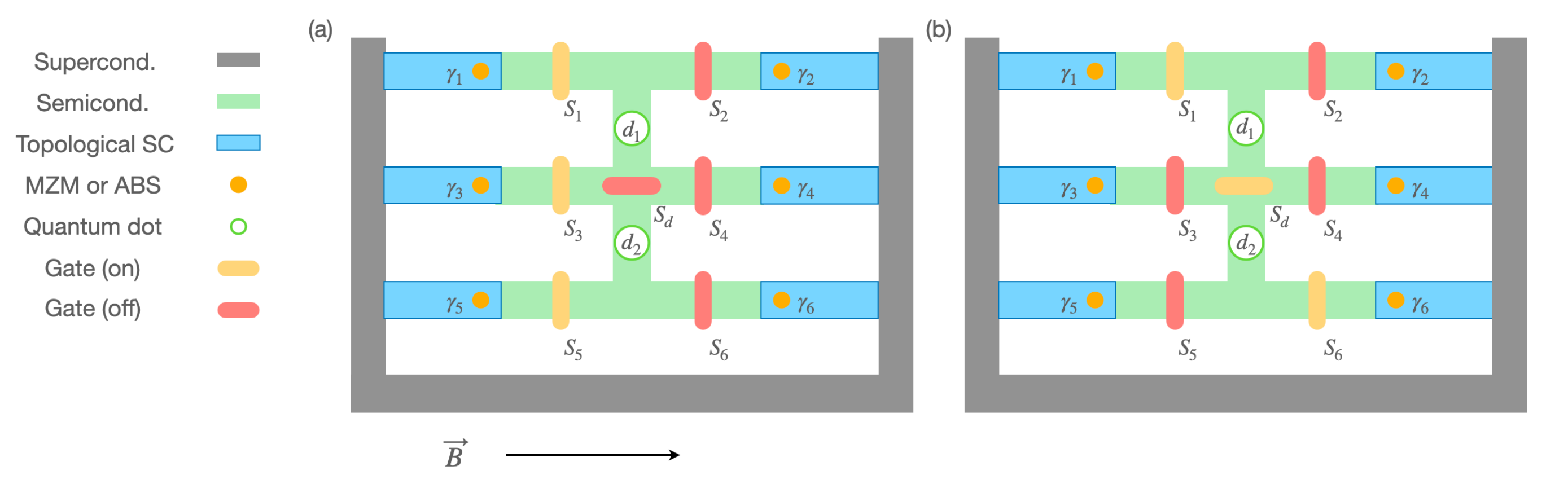}
    \caption{\label{fig:setup1} The example design of six MZMs. The
    blue wires are topological superconductor wires, which consists a wire of topological semiconductor on a superconductor base put in a magnetic field along the wire \cite{Das2012} (an example is InAs on top of Al). 
    And the gray wires are superconductor wires that connect all the wires as an island.
    The green wires are semiconductor that also connect the wires together in the middle.
    The tunneling strength $t_i$ between the MZMs and the QDs can be depicted as gates.
    Each tunneling can be switched on and off by adjusting the gate voltage.
    Figure $(a)$ shows the status of the gates when single-site measurement on site $1$ by QD $d_1$ is performed.
    The first step of our protocol would be performing such measurements to every site in one subsystem to eliminate the bipartite entanglement.
    And figure $(b)$ shows the paired measurement across two subsystems on site $1$ and site $6$.
    Here by fully opening the tunneling between $d_1$ and $d_2$, the two QDs acts as one QD coupled with site $1$ and site $6$.
    The parity measurements consist of such bipartite measurements, and the witness is calculated based on the outcome of such measurements.
    Note that it is not drawn to scale. 
    In reality the length of the TS wires $L_W$ is much longer than in this simple illustration to make $L_W\gg\xi$, so the ground state degeneracy of the MZM is guaranteed by $L_W \gg \xi$.}
\end{figure*}

Consider, for instance, a scenario involving six one-dimensional topological superconductors. This six-site arrangement is illustrated in Figure~\ref{fig:setup1}, featuring an array of topological superconductors linked to a superconductor at one terminus, while poised to accommodate multiple MZM pairs at the opposite end. Through the adjustment of an external magnetic field, MZMs are induced as the lowest-energy excitations at the extremities of the setup. Our discussion will delve into a detailed examination of our protocol based on this simplified configuration. The distribution of three MZM pairs within this six-site framework is depicted in Figure~\ref{fig:sites}, where a specific MZM pairing configuration has been chosen, namely pairs $1-5$, $3-6$, and $2-4$.

In this context, a possible state of the MZM-paired system, characterized by an odd total parity, is expressed as:
\begin{equation}\label{mzmState}
\begin{aligned}
    |\Phi\rangle_{\text{MZM}} =& A|0_{15}1_{36}0_{24}\rangle + B |0_{15}0_{36}1_{24}\rangle \\
    &+C |1_{15}1_{36}1_{24}\rangle +D |1_{15}0_{36}0_{24}\rangle
\end{aligned}
\end{equation}
Herein, the subscript specifies the pair's spatial positioning, while the binary indicators $0$ or $1$ denote the even or odd MZM parity across the two sites, respectively, diverging from the fermion number attributed to a singular fermionic state.

In the framework of this model, the system is divided into two distinct segments: $1, 2, 3$ and $4, 5, 6$, as delineated in Figure~\ref{fig:sites}. In particular, within the three specified pairings, at least one pairing invariably bridges these two subsystems, with the pairing of $3-6$ serving as a case in point in the current context. This specific configuration engenders nonlocal correlations among the MZMs. Consequently, states resulting from such MZM pairings manifest as inseparable under the aforementioned partitioning. In essence, the composite state $|\Phi \rangle_{\text{MZM}}$ cannot be decomposed into a simple tensor product of states corresponding to the two subsystems, such as $|\Phi \rangle_{135} \otimes |\Phi\rangle_{246}$. This characteristic underscores the entangled nature of MZM-paired states, highlighting their intrinsic nonlocality and the impossibility of expressing them as product states of the divided system.

\subsection{Local Andreev bound modes}\label{sec:ABS}
At the boundaries of a superconductor, an injected electron can undergo a process where it is reflected back as a hole, while simultaneously forming a Cooper pair within the superconductor.
This phenomenon, known as Andreev reflection, effectively converts an electron-like excitation on the normal side into a hole-like excitation, conserving charge across the interface. Andreev reflection is also responsible for the formation of Andreev bound states (ABSs) when a semiconductor is coupled to superconducting nanowires. ABSs are localized states that can appear at the interface between the superconductor and the semiconductor. 

Unlike MZMs that are predicted to exhibit non-abelian statistics and nonlocal properties, ABSs do not inherently possess these characteristics. Both ABSs and MZMs can manifest experimentally as zero-bias conductance peaks in tunneling experiments, which complicates the distinction between these two types of states. 
However, ABSs are typically characterized by localized excitations that do not influence the non-local properties of the system's many-body ground state, contrasting with the inherently non-local nature of many-body ground states associated with MZMs. MZMs are theoretically predicted to exist in topological superconductors and are expected to contribute to non-local quantum correlation across the system, a property not shared by ABSs. This non-local correlation underpins the interest in MZMs for topological quantum computing, where the spatial separation of MZMs protects the system from local perturbations and decoherence, unlike the localized states of ABSs.
Distinguishing between MZMs and ABSs in experimental setups is challenging, as both can produce similar signals in conductance measurements. On the other hand, failing to carefully examine MZMs and their nonlocal nature could lead to fundamentally different outcomes when applying these devices in quantum information processing. 

The ABS operators can be represented in terms of local fermionic operators as follows:
\begin{equation}\label{eq:ABS_state}
    c_i^{\text{ABS}} = |u_i| c_i + |v_i| c_i^\dagger.
\end{equation}
In this representation, \(c_i\) and \(c_i^\dagger\) denote the fermion annihilation and creation operators at site \(i\), respectively, while \(|u_i|\) and \(|v_i|\) are coefficients that describe the particle and hole components of the ABS. When \(|v_i| = 0\), the ABS is purely electron-like, reducing to a local fermionic state.
As a specific instance of an ABS, the local fermionic state will be explored in detail later in Appendix~\ref{app:AWCandidate}. Expressing ABS creation and annihilation operators in terms of fermionic operators simplifies the analysis of their parity properties and aids in the calculation of witness operators for identifying topological states. Therefore, this fermionic representation will be used consistently throughout our discussion.




At each site, the local ABS parity can be characterized as either odd (with an odd number of fermions, denoted as \(|1\rangle\)) or even (with an even number of fermions, denoted as \(|0\rangle\)). An Andreev bound state exhibiting an odd total fermion number in a 6-site model can be expressed as:
\begin{equation}\label{ABSstate}
    |\Psi \rangle_{\text{ABS}} = A|000001\rangle + B|000010\rangle + \cdots
\end{equation}
It is important to note that each basis state is the product of localized states, indicating the absence of non-local correlations typically induced by topological pairing mechanisms. However, such states can exhibit entanglement between two subsystems, for example, \((1,3,5)\) and \((2,4,6)\) for the six-site toy model in Fig.~\ref{fig:sites}. When considering the projection onto these subsystems, the overall six-site ABS configuration becomes separable, represented as \(|\Psi \rangle_{135} \otimes |\Psi\rangle_{246}\), illustrating the localized nature of the specific subsystems (i.e. ABSs).



On the other hand, due to the intrinsic nonlocal pairing characteristic of Majorana zero modes (MZMs), projection operations across the boundary of a system harboring MZMs do not result in a separable state. This distinct behavior underlines a fundamental difference between MZM-paired states and ABSs. The persistence of nonlocal entanglement in MZM pairs, even after such projection operations, is a testament to their topological nature. Consequently, a projected MZM state retains its nonlocality, as shown below:
\begin{equation}\label{mzmOdd}
\begin{aligned}
    |\Phi_{\text{odd}} \rangle = & A'|0_{15}1_{36}0_{24}\rangle + B' |0_{15}0_{36}1_{24}\rangle \\
    & + C' |1_{15}1_{36}1_{24}\rangle + D' |1_{15}0_{36}0_{24}\rangle,
\end{aligned}
\end{equation}
where the coefficients \(A'\), \(B'\), \(C'\), and \(D'\) are transformations of the original coefficients \(A\), \(B\), \(C\), and \(D\), respectively. Notably, the subscript notation, which indicates site-specific pairings, is preserved post-projection to reflect that the operation does not alter the fundamental pairing mechanism or parity of the system. Henceforth, for simplicity and without loss of generality, we will continue to use this pairing notation for the six sites, omitting the subscript that designates specific MZM pairs.



Therefore, leveraging the property that MZM pairing cannot be erased by local projection operations (see Section~\ref{sec:protocol} for more details), we can design a witness to distinguish between ABSs and MZMs.

\subsection{\label{sec:tunn}The tunneling physics}

To perform the MZM witness protocol, parity measurements between any two sites are essential. Previous studies, such as Romito and coworkers~\cite{romito2017ubiquitous}, demonstrated the measurement of six terminal MZM states using four quantum dots (QDs). Drawing inspiration from scalable architectures proposed in ~\cite{Karzig2017Scalabledesigns} and the Majorana box qubit configuration delineated in ~\cite{Plugge2017}, we introduce a viable and scalable experimental arrangement for a six-site model. In our design, six MZM sites are symmetrically placed, flanking two centrally located QDs, as depicted in Figure~\ref{fig:setup1}. The main features of our experimental framework are summarized as follows.

\begin{enumerate}
\item In our setup, there are several topological superconductor (TS) wires hosting Majorana Zero Modes (MZMs) at their ends. Each wire, a composite of a spin-orbit coupled semiconductor and a superconducting layer, enters a TS phase under an external magnetic field. These six wires are interconnected at one end by a superconducting bulk, enabling the potential formation of MZMs at their free ends. This design harnesses the interplay between semiconductors and superconductivity to facilitate the conditions necessary for MZMs.
\item It should be noted that the TS wires are parallel to each other.
The design of the full T-junction structure for TS illustrated in~\cite{aasen2016milestones} is avoided to ensure that the external magnetic field aligns with all the wires. This arrangement facilitates the simultaneous induction of a TS phase across all wires through a strategically oriented magnetic field.
\item The TS wires are deliberately fabricated to be lengthy, such that \(L_W \gg \xi\), where \(\xi\) represents the coherence length within the TS framework. This design is important as it significantly reduce the ground state degeneracy splitting, attributable to the overlap of the MZM wave functions, and is quantitatively expressed as \(\sim \exp(-L_W/\xi)\) \cite{Das2012}. 
\item The charging energy, denoted by \(E_C\), is instrumental in preventing quasiparticle events and exponentially safeguarding the parity of MZMs, following \(\exp(-E_C/T)\). However, an increase in the length of the TS wires (\(L_W\)) inversely impacts \(E_C\), diminishing the charging energy. Therefore, it is crucial to establish an optimal balance between \(E_C\) and \(L_W\). This relationship further justifies the examination of quasiparticle poisoning, as elaborated in Section~\ref{sec:MZMerror}.
\item According to \cite{Karzig2017Scalabledesigns}, the structure of the six MZMs forms a hexon design of a qubit.
Moreover, with the help of the two QDs in the middle, all projective measurements over a certain subset of the six sites are possible.
Each point in the center of two MZMs has two gates connected to the neighboring MZM sites, and the two QDs are also coupled together by a gate.
By tuning the voltage of the gates, the tunneling between the QDs and the MZM island, as well as the level of the QDs, can be adjusted. Thus, the interaction between the MZM sites and the QDs can be switched on or off by the gates.
The status of the gates for a two-site parity measurement is shown in the right panel of Figure.~\ref{fig:setup1}.
\end{enumerate}





According to \cite{Karzig2017Scalabledesigns}, the Hamiltonian of the system considered here (showed in Figure.~\ref{fig:setup1}) has three parts
\begin{equation}\label{eq:totalH}
    H = H_0 + H_{QD} + H_{tunn}.
\end{equation}
The first term is the Hamiltonian of the island, which consists of superconductor wires and the semiconductor base.
The second term refers to the QDs coupled to the ends of the nanowires.
The third term is the tunneling Hamiltonian between the QDs and the states at the ends of the wires.

The island Hamiltonian $H_0$ in Eq.(\ref{eq:totalH}) can be written in two parts: the BCS Hamiltonian $H_{\text{BCS}}$ and the charging energy $H_{\text{C}}$
\begin{equation}
    H_0 = H_{\text{BCS}} + H_{\text{C}} = H_{\text{BCS}} + E_C (N_s - N_g).
\end{equation}
Here $E_C$ is the charging energy, $N_s$ is the charge number on the island, and the charge $N_g$ effectively represents the external electric field.
As in \cite{Karzig2017Scalabledesigns} the second term in Eq.(\ref{eq:totalH}) is the effective Hamiltonian of QDs
\begin{equation}
    H_{\text{QD}} = h\hat{n}_f + \epsilon_C(\hat{n}_f - n_g)^2.
\end{equation}
Here we ignore the spin of the QD because of the strong magnetic field~\cite{LutchynPRL10,Oreg_helical_2010}.
The first term here is the orbital energy with coefficient $h$, and the second term is the charging energy. $\hat{n}_f$ is the occupation number of the QD. $n_g$ also effectively represents the external electric field on the QD.
$\epsilon_C$ is the charging energy of the QD.
Comparing with the superconducting charging energy $E_C$, $\epsilon_C$ is much larger, i.e. $\epsilon_C \gg E_C$.
So the two lowest energy state of the system can be labeled by $|n_f = 0, 1\rangle$.
The QD energy of occupancy $0$ and $1$ are
\begin{equation}
    \begin{aligned}
        \epsilon_0 &= \epsilon_C n_g^2,\\
        \epsilon_1 &= \epsilon_C (1-n_g)^2 + h.
    \end{aligned}
\end{equation}

The third term in Eq.(\ref{eq:totalH}) is the key to MZM parity measurements.
The parity-dependent shift of ground state energy is induced by the tunneling between the island and the QD.
The shift can be calculated following the perturbation method in \cite{Karzig2017Scalabledesigns}.

With QD generating operators represented by $f_1^\dagger, f_2^\dagger$, we can measure the 2-site parity, for example, the parity between site 1 and site 2. The tunneling Hamiltonian would be
\begin{equation}\label{eq:MZMtunn}
    H_{tunn,MZM}^{(1,2)} = -\mathrm{i}\frac{e^{-\mathrm{i}\phi/2}}{2}\left( t_1 f_1^{\dagger} \gamma_1+ t_2 f_1^{\dagger} \gamma_2 \right)+\text { H.c. }
\end{equation}
$e^{\mathrm{i}\phi/2}$ is the phase shift when adding an electron to the island, $e^{\mathrm{i}\phi/2}|N_s\rangle = |N_s +1\rangle$.
$t_1$ and $t_2$ are the coupling strength between the QD and site 1 and 2.
Since the device is placed in a strong magnetic field to produce topological superconductivity, $t_1$ and $t_2$ are both complex number.
Considering the perturbation induced by the tunneling Hamiltonian Eq.(\ref{eq:MZMtunn}), one can get~\cite{Karzig2017Scalabledesigns}
\begin{equation}\label{eq:MZM_E}
    \begin{aligned}
        \varepsilon_1^{\mathrm{MZM}}&=E_C N_g^2+\epsilon_1-\frac{\left|t_1\right|^2+\left|t_2\right|^2+\mathrm{i} p_{12}\left(t_1^* t_2-t_1 t_2^*\right)}{4\left[E_C\left(1-2 N_g\right)+\epsilon_0-\epsilon_1\right]}\\
        \varepsilon_0^{\mathrm{MZM}}&=E_C N_g^2+\epsilon_0-\frac{\left|t_1\right|^2+\left|t_2\right|^2-\mathrm{i} p_{12}\left(t_1^* t_2-t_1 t_2^*\right)}{4\left[E_C\left(1+2 N_g\right)+\epsilon_1-\epsilon_0\right]}
    \end{aligned}
\end{equation}

This tunneling Hamiltonian is for the MZM case.
Next, we want to calculate the ground-state energy splitting given by ABS under the same settings.
As presented in Eq.(\ref{eq:ABS_state}), the ABS operator is a superposition of electrons and holes.
We would first like to consider a simpler situation, where the coefficient of the hole is 0. 
The ABS creation and annihilation operator is now equivalent to a fermion creation and annihilation operator.

For fermionic state, the operators $\gamma_i$ is then replaced by the ordinary fermionic operators $c_i$
\begin{equation}\label{eq:Ftunn}
    H_{tunn,F}^{(1,2)} = -\frac{\mathrm{i}e^{-\mathrm{i}\phi/2}}{2}\left(t_1 f_1^{\dagger} c_1+ t_2 f_1^{\dagger} c_2 \right)+\text { H.c. }
\end{equation}

The lowest energy perturbation of the fermionic state gives

\begin{equation}\label{eq:F_E}
    \begin{aligned}
        \varepsilon_1^F & =E_C N_g^2+\epsilon_1-\frac{\left|t_1\right|^2+\left|t_2\right|^2-\left(\left|t_1\right|^2+\left|t_2\right|^2\right) b_{12}^{\dagger} b_{12}}{4\left[E_C\left(1-2 N_g\right)+\epsilon_0-\epsilon_1\right]} \\
        \varepsilon_0^F & =E_C N_g^2+\epsilon_0-\frac{\left|t_1\right|^2+\left|t_2\right|^2+\left(\left|t_1\right|^2+\left|t_2\right|^2\right) b_{12}^{\dagger} b_{12}}{4\left[E_C\left(1+2 N_g\right)+\epsilon_1-\epsilon_0\right]}
    \end{aligned}
\end{equation}

Here, $b_{12}^\dagger  \in \{b_{ij}^\dagger\}$ is the unitary transformation of the two fermionic operators.
\begin{equation}\label{eq:b_F}
    b_{12}^\dagger = \frac{t_1^* c_1^\dagger + t_2^* c_2^\dagger}{\sqrt{\left|t_1\right|^2+\left|t_2\right|^2}}
\end{equation}

The energy gap depends on the expectation value of the effective operator $b_{12}^\dagger b_{12}$.
Then for the fermionic state, measuring the energy gap in the same way gives the measurement value of $b_{12}^\dagger b_{12}$.

More generally, for the Andreev bound states, the tunneling Hamiltonian is
\begin{equation}\label{eq:AbsTunn}
\begin{aligned}
    H_{tunn,ABS}^{(1,2)} = -\frac{\mathrm{i}e^{-\mathrm{i}\phi/2}}{2}[&t_1 f_1^{\dagger} (\left| u_1\right| c_1 + \left| v_1\right| c_1^\dagger)\\
    &+ t_2 f_1^{\dagger} (\left| u_2\right| c_2 + \left| v_2\right| c_2^\dagger)]+\text { H.c. }
\end{aligned}
\end{equation}
The difference between Eq.~(\ref{eq:AbsTunn}) and the Eq.~(\ref{eq:Ftunn}) is that the QD interacts with both the hole and particle on the island.
The interaction is normalized, that is, $\left| u_i\right|^2 + \left| v_i\right|^2 = 1$.
For the ABS case, we can still get an effective measurement operator $b^\dagger_{12,ABS} b_{12,ABS}$, here the effective operator 
\begin{equation}\label{eq:abs_b}
    b_{12,ABS}^\dagger = \frac{t_1^* (\left| u_1\right| c_1^\dagger + \left| v_1\right| c_1) + t_2^* (\left| u_2\right| c_2^\dagger + \left| v_2\right| c_2)}{\sqrt{\left|t_1\right|^2+\left|t_2\right|^2}}.
\end{equation}
The detailed calculation of the perturbation is in Appendix~\ref{app:Perturbation}.

\section{\label{sec:witness} Entanglement witness}

From the previous sections, we understand that the fundamental distinction between MZMs and ABSs is grounded in their separability and the nonlocal versus local nature of their information encoding. A variety of techniques are available for detecting and quantifying the bipartite entanglement in general quantum systems, with several popular methods reviewed in \cite{Horodecki09}. 

Regarding the special case of MZMs, in the realm of nonlocal pairing-induced correlations of MZMs, the violation of the Bell and CHSH inequalities has been explored, as detailed in \cite{romito2017ubiquitous}.
Beyond these inequalities, a broader approach for differentiating entangled states from separable ones involves the use of entanglement witnesses for general quantum systems. In this study, we employ a specific type of witness operators to develop a protocol for detecting Majorana non-local behaviors.

Our witness is inspired by the witness defined in \cite{Yu05}.
In \cite{Yu05}, the authors define a witness operator consisting of the direct product of complete sets of local orthogonal operators.
They prove that such a witness operator can detect bound entangled states.
However, such a witness operator is not directly applicable to the task of distinguishing between MZM paired states and ABS, because entangled states and non-entangled states exist in the same dimensional space, while MZM paired states and ABS do not.

\subsection{Definition}

For a bipartite system described by the state $\rho_{AB}$, the system is considered separable if it can be expressed as the product of the states of its subsystems, $\rho_A \otimes \rho_B$. Assuming that subsystems $A$ and $B$ have identical dimensions, the criteria for an operator $W$ to qualify as an entanglement witness are as follows:
\begin{enumerate}
    \item It must yield nonnegative mean values when applied to all separable states, i.e., 
    \begin{equation}\label{eq:candidate}
        \left\langle\psi_{A}\left|\left\langle\phi_{B}\right|W\left| \psi_{A}\right\rangle\right| \phi_{B}\right\rangle \geqslant 0
    \end{equation}

    \item It must possess at least one negative eigenvalue, indicating its ability to detect entanglement.
\end{enumerate}
If an operator $W$ satisfies only the first condition, it is classified as an \textbf{entanglement witness (EW) candidate}. \cite{romito2017ubiquitous}





\begin{figure}[htpb]
    \centering
    \includegraphics[width=0.9\linewidth]{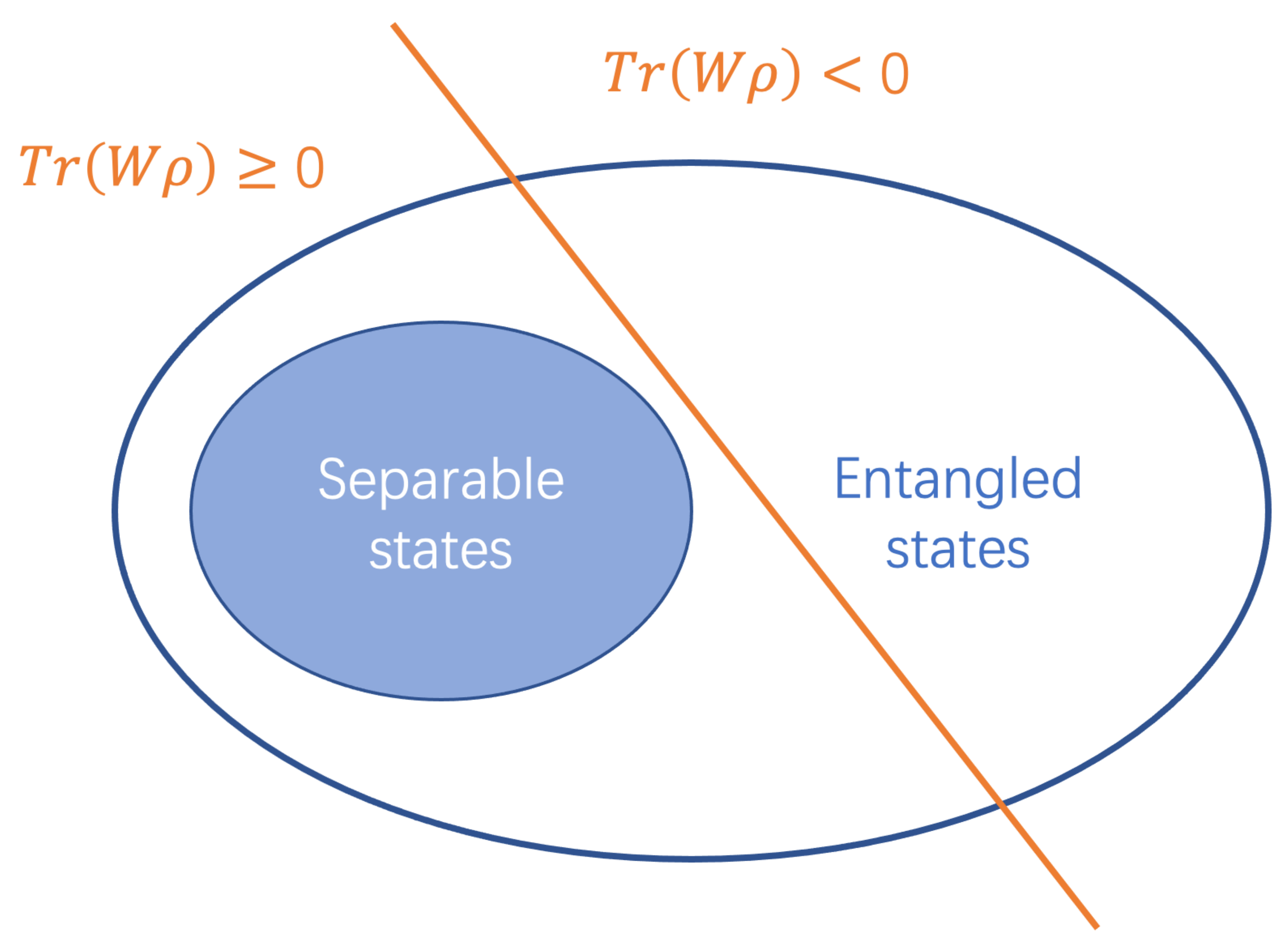}
    \caption{\label{fig:witness}The graphic illustration of the definition of an entanglement witness. The big eclipse represents the whole Hilbert space, while the shadowed part is the separable subspace.\cite{Horodecki09}}
\end{figure}

In other words, a necessary condition for the separability of a state $\rho_{AB}$ is:
$\text{Tr}(W \rho_{AB}) \geq 0$, and a sufficient condition for $\rho_{AB}$ to be an entangled state is: $\text{Tr}(W \rho_{AB}) < 0$. Specifically, the trace operation applied to the product of the entanglement witness $W$ and the state $\rho_{AB}$ is described by:
\begin{equation}\label{eq:EWCondition}
    \operatorname{Tr}(W \rho) 
    \begin{cases}
        \geq 0 & \text { separable or entangled} \\ 
        <0 & \text { entangled }
    \end{cases}
\end{equation}
This indicates that the trace value of the witness $W$ is negative solely within the entangled subspace. However, within the non-negative subspace of $W$, the state could be either separable or entangled.
The characteristics of an EW are illustrated in Figure~\ref{fig:witness}.
For any entangled state, there exists an EW that can detect it, a property known as the completeness of the EW.
Furthermore, the effectiveness of an EW increases with the number of entangled states it can detect.
The search for the optimal EW corresponds to an optimization problem, with completeness and optimization of entanglement witnesses discussed in Appendix~\ref{app:EW}.

To distinguish between MZMs and ABS is a binary classification problem.
In this context, the measured value of a witness operator serves as the classifier. 
The constructed witness is based on nonlocal measurements of bound state pairs, such as MZM pairs or ABS pairs, within the system. 
For two different types of physical state, the same measurement procedure has different forms due to different physical structures.
The two forms of the witness are denoted as the Andreev bound state witness $W_{A}$ (AW) and the Majorana zero modes witness $W_{M}$ (MW).
The final value of $\operatorname{Tr}(W\rho)$ is calculated from the measured value regardless of the nature of $\rho$.

If the witness for ABSs, \( W_A \), qualifies as \textbf{an EW candidate}—meaning that for all separable states \( \rho \), \( \operatorname{Tr}(W_A \rho) \geq 0 \)—the false positive rate of the classifier would be \( 0\% \).
In contrast, the witness for MZM-paired states, \( W_M \), needs to achieve values that are as negative as possible. The higher the fraction of the state space satisfying \( \operatorname{Tr}(W_M \rho) < 0 \), the higher the true positive rate of the classification.


\subsection{CHSH Entanglement Witness}
Well-known entanglement witnesses, such as the swap operator for two subsystems, have been extensively studied in the literature \cite{horodecki1997separability,Barbieri03,Bourennane04,Toth05,Yu05,Eisert_2007,Horodecki09,Sperling13}.
As an example, we can write an entanglement witness corresponding to the CHSH inequality in \cite{CHSH1969}.
The discussion in \cite{romito2017ubiquitous} on the violation of the CHSH inequality of MZM pairs leads to the CHSH entanglement witness.

The CHSH inequality is obeyed by all separable bipartite states,
\begin{equation}
    \begin{aligned}
        \mathcal{C} \equiv\left|\left\langle\hat{L}_1 \hat{R}_1\right\rangle-\left\langle\hat{L}_1 \hat{R}_2\right\rangle\right|&+\left|\left\langle\hat{L}_2 \hat{R}_2\right\rangle+\left\langle\hat{L}_2 \hat{R}_1\right\rangle\right| \leq 1,\\
    &\text{ for all separable states.}
    \end{aligned}\label{eq:CHSH}
\end{equation}
where $\hat{L}_1$, $\hat{L}_2$, $\hat{R}_1$, and $\hat{R}_2$ are local operators for the left and right subsystems.
If the subsystems have dimension of $2$, then the local operations can be written as the Pauli operations on a spin, $\hat{\mathbf{\sigma}}\cdot \mathbf{n}$. The Pauli basis is $\hat{\mathbf{\sigma}} = (\sigma_x, \sigma_y, \sigma_z)/\sqrt 2$.
And $\mathbf{n} = (\sin \theta_L \cos \phi_L, \sin \theta_L \sin \phi_L, \cos \theta_L)$ is a unit vector.
The CHSH inequality can be violated by entangled bipartite states.
Follow this inequality, we can construct an operator $W$ that satisfies the entanglement witness conditions:
\begin{equation}
    W_{CHSH} = I - \sum_{i,j \in \{1,2\}}(-1)^{k(i,j)}\left\langle\hat{L}_i \hat{R}_j\right\rangle \label{eq:CHSHEW}
\end{equation}
To measure the CHSH witness, the measurement needed for the witness are $\left\langle\hat{L}_i \hat{R}_j\right\rangle$, $i, j \in \{1,2\}$.
For all separable states, the CHSH inequality is satisfied and $W \geq 0$. 
$W < 0$ corresponding to the violation of the CHSH inequality, it can therefore be concluded that the state is an entangled state.

Now we take the six-MZM bipartite system as an example, if the MZM labeled 1, 3, and 5 are the left subsystem, the possible operators on $L$ has the form of:
\begin{equation}
\begin{aligned}
    \hat{L}
    = -\frac{i}{2}(\cos \theta_L \gamma_1 \gamma_3 &+ \sin \theta_L \cos \phi_L \gamma_3 \gamma_5\\
    & +\sin \theta_L \sin \phi_L \gamma_5 \gamma_1),
\end{aligned}
\end{equation}
with the Pauli operators on the left subsystem $\{I,-i\gamma_3 \gamma_5, -i \gamma_5 \gamma_1, -i \gamma_1 \gamma_3\}/\sqrt{2}$. And similarly for the right counterpart the Pauli operators are $\{I,-i\gamma_4 \gamma_6, -i \gamma_6 \gamma_2, -i \gamma_2 \gamma_4\}/\sqrt{2}$.
The Pauli operators are MZM parity measurements, which can be measured by the tunneling method described in Section~\ref{sec:tunn}.

However, due to the different physical nature of MZMs and ABSs, the same measurement operations yield distinct outcomes. In the case of local ABSs, the dimension of the relevant subspace is not $2$. Consequently, the previously discussed Pauli measurements used in the CHSH witness are incompatible with the ABS subspace. As a result, the CHSH witness cannot be applied to this classification task. To address this limitation, we propose a new witness operator specifically designed for this purpose.

\subsection{\label{sec:MW}The MZM witness (MW)}
For the MZM case, the witness operator consists of MZM parity measurements $i\gamma_i \gamma_j$, as showed in Eq.~(\ref{eq:MZM_E}). So the EW can be constructed as
\begin{equation}\label{eq:MZMWitness}
    W_M = I - \sum_{\langle i,j \rangle} a_{ij} \left \langle \mathrm{i}\gamma_i \gamma_j\right \rangle = I - \sum_{\langle i,j \rangle} a_{ij} p_{ij}.
\end{equation}
Here the pairs $\langle i,j \rangle$ are all nonlocal pairs, i.e. pairs that $i$ belongs to one subsystem and $j$ belongs to the other.
This is because we only need nonlocal information for the task.

For this witness operator, we only need it to have a negative eigenvalue, that is, there exists a state $\rho$ and its neighborhood $\rho_{MZM}$such that $\operatorname{Tr}[\rho_{MZM} W] <0$.
Since we do not have a priori knowledge of the state, the set of parameters $\{a_{ij}\}$ is the same as the ABS witness.
Thus the restriction on $\{a_{ij}\}$ is that the ABS witness $W_A$ is an EW candidate.
We will give this constraint in the next subsection, and prove that under this constraint the MW has negative value.

\subsection{\label{sec:AW}The Andreev bound state witness (AW)}
For the case of Andreev bound state, the tunneling of particle and hole on each site can have different strength.

The operators measured in the witness would be in terms of $b_{ij,ABS}^\dagger b_{ij,ABS}$, which is showed in Eq.~(\ref{eq:abs_b})

\begin{equation}
    W_A = I - \sum_{\langle i,j \rangle} a_{ij} \left\langle b_{ij,ABS}^\dagger b_{ij,ABS}\right\rangle.
\end{equation}

Following the calculation in Appendix~\ref{app:AWCandidate}, 
the EW candidate condition for the AW is:
\begin{equation}
     \operatorname{Tr} [W_A (\rho_1 \otimes \rho_2)] \geq 1- \sum_{ij}a_{ij} (1+\sqrt{2}+\frac{2|t_i^* t_j|}{\left|t_i\right|^2+\left|t_j\right|^2}) \geq 0.
\end{equation}
The constraint of $\{a_{ij}\}$ is related to the tunneling coefficient $t_i$ and $t_j$.
Take $T_{ij} \equiv \frac{2|t_i^* t_j|}{\left|t_i\right|^2+\left|t_j\right|^2}$,
for all complex $t_i$ and $t_j$, $0\leq T_{ij}\leq 1$.

\begin{figure*}[tb]
    \centering
    \includegraphics[width=\linewidth]{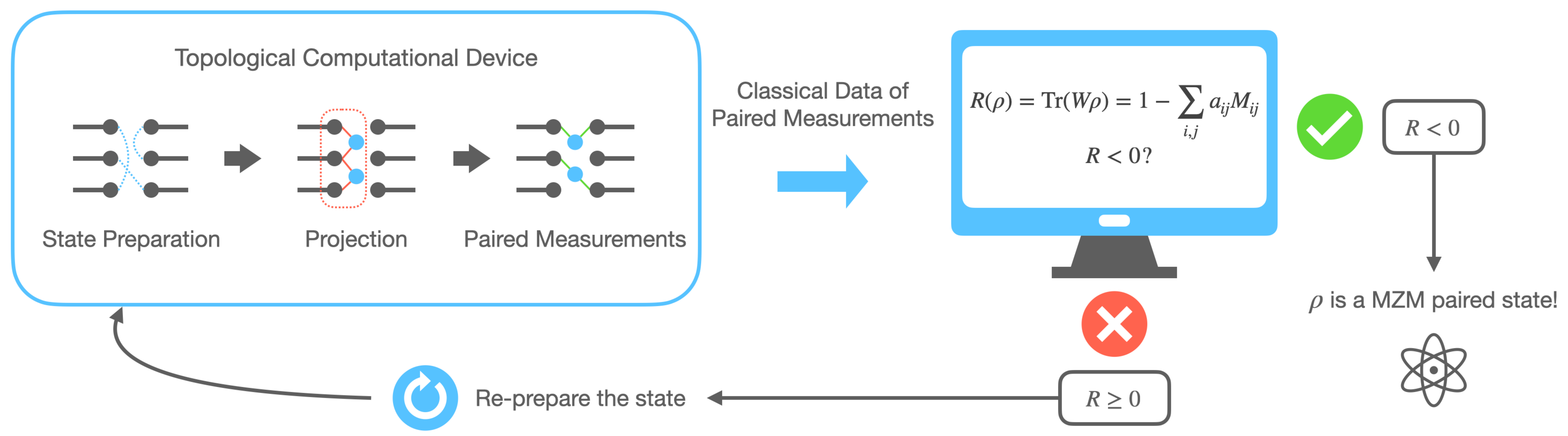}
    \caption{\label{fig:protocol} The graphical representation of the MZM detection protocol. The input of the protocol is the prepared initial state and its copies.
    After one cycle, depending on the witness result, the protocol either outputs a negative value, thus verifies an MZM paired state, or outputs a non-negative value, thus gives no valid solution and goes back to the start to repeat-- for another prepared state.}
\end{figure*}

\section{\label{sec:protocol}Protocol}
In this section, we will give a detailed protocol on how to use the parity witness to distinguish MZM states from ABS states.
The protocol is designed to make sure that the output state of the protocol definitely has MZM pairs.
The operation needed for our protocol is the projection for elimination of entanglements, and the paired measurements of all pairs cross two subsystems.
Quantum dots (QDs) are used to couple MZM sites to detect nonlocality in the previous study\cite{romito2017ubiquitous}.
In our protocol, we also utilize the QDs coupled with the MZM sites to accomplish the required operations.

The whole protocol is sketched in Figure~\ref{fig:protocol}.
Note that we use the example of the 6-site model in Figure~\ref{fig:setup1} for the demonstration of the protocol.
In this section, we will present the protocol step by step: 
the projection, the parity measurement, and the calculation of the trace value.

\subsection{Entanglement elimination}
Before employing witness measurements to confirm MZM pairing, we must first disentangle the two subsystems, eliminating both bipartite quantum entanglement and classical correlation from ABS.
For MZM paired states, the topological pairing persists after entanglement elimination.
Therefore, any nonlocal correlations detected by the witness must arise from the topological nature of MZMs.

Let us first examine the case of ABS.
In the experimental setup shown in Figure~\ref{fig:setup1}, the gates can be adjusted to sever the connection between each site.
By controlling gate voltages, the quantum dot can couple exclusively with one neighboring site, as shown in Figure~\ref{fig:setup1}(a).
When ABS exhibits initial entanglement between subsystems, this quantum correlation becomes highly susceptible to noise, particularly in solid-state systems with no interaction terms in its Hamiltonian.
Without careful state preparation (such as robust quantum codes or noise-resistant encoding), the entanglement decays rapidly due to decoherence and environmental interactions.
In~\cite{yu_finite-time_2004}, the authors demonstrate that under pure vacuum noise, two entangled qubits completely disentangle within a finite time.
Experiments have determined the lifetime of several types of superconducting qubits, as detailed in a recent review paper~\cite{siddiqi_engineering_2021}.
Typically, without deliberate and careful preparation, the two parts of the ABS system are not entangled. In the worst-case scenario, the bipartite ABS system will generally disentangle over a short timescale, characteristic of solid-state systems.

Now we consider the MZM case.
MZMs are characterized by a nonlocal basis (as evident from the Majorana operator Eq.~(\ref{eq:MZM_op})), and local or single-site projection measurements do not alter their fundamental nonlocal structure.
With all site-wise interactions closed by the gates, the local operations on site $i$ can only include $\gamma_i$ terms.
Since $\gamma_i \gamma_i^\dagger$ is trivial, operations are limited to linear terms of $\gamma_i$, which preserve the nonlocal correlations arising from MZM pairing.

Consequently, after local operations, while ABS states lose their bipartite correlation, MZM paired states maintain their nonlocal paired form as shown in Eq.~(\ref{mzmState}).

\subsection{Paired measurement}
Parity measurements are conducted on the projected states obtained from the previous step to investigate their nonlocal properties. Specifically, we measure the two-point fermion parity for all possible pairs spanning the two subsystems, as these measurements are designed to probe the nonlocality inherent in the state. In the illustrative example, the pairs to be measured include $\{1,2\},\{1,4\},\{1,6\}, \dots ,\{5,6\}$.

The parity measurement of two MZMs can be conducted on devices by interference parity measurements with high accuracy, as shown recently by Microsoft in \cite{aghaee_interferometric_2025}.
The Hamiltonian model as well as the perturbation analysis used by Microsoft group is similar to the model discussed in our work as in Section~\ref{sec:tunn}.
Nevertheless, the physical quantity directly measured experimentally is the quantum conductivity of the quantum dot instead of the parity.
To process the experimental data, the relation between the quantum conductance and the parity is calculated by open system dynamics to take into account the influence of an additional readout circuit.
It is worth mentioning that the author's analysis of the quasi-MZM scenario case indicates that merely using parity measurement cannot distinguish between such topological trivial states and MZM pairs.

In our toy model, we take a simplified measurement representation.
Although the measurement conducted on a real device requires three QDs, there is only one effective QD with two effective coupling parameters in the physical model.
The effective coupling strength of the QD to each end of the nano wire, $t_L$ and $t_R$, can be tuned by the flux in the circuit as well as the gate voltages of each physical QDs.
Thus for a two-site parity measurement, we only 
show one QD that couples the relevant sites, and the two coupling strength is controllable.
The Hamiltonian of such model is shown in Eq.~\ref{eq:MZMtunn}.
The parity measurement of sites 1 and 2 is shown in Figure~\ref{fig:setup1}.
This simplified device model is compatible with the circuits in Microsoft parity measurement.
Such a device model, i.e. the tetron-based device, also appeared in Microsoft's recent topological quantum computing roadmap~\cite{aasen_roadmap_2025} and earlier works~\cite{Karzig2017Scalabledesigns}.

In a multi-Majorana device, by tuning the gate voltage, one can controllably open or close the coupling between each site and QD, thereby determining which pairs of sites are measured.
The measurement procedure itself is identical for both ABSs and MZMs, differing only in the specific quantity being measured (cf.\ Section~\ref{sec:AW} and Section~\ref{sec:MW}).
Because these paired measurements generally do not commute, each measurement must be carried out on a freshly prepared projected state. Once a parity measurement has been performed, the state is discarded and replaced by a new projected state for the next measurement.

In summary, the parity measurement procedure yields the expectation values for all possible paired measurements spanning the two subsystems. These results encode the nonlocal information present in the state. The collected data serve as the foundation for the subsequent analysis, where the nature of the tested state—whether it corresponds to MZMs or ABSs—will be determined based on its nonlocal characteristics.

\subsection{Calculating the witness value}

Using the paired measurement data obtained in the previous step, the value of $\operatorname{Tr}(W \rho)$, where $\rho$ denotes the projected state, can be computed. This computation is performed by a classical computer that processes the data received from the experimental system.

In the witness operator $W$, the parameters $\{a_{ij}\}$ are selected to satisfy the witness condition outlined in Section~\ref{sec:witness}. As previously discussed, the parameters for both the ABS witness (AW) and the MZM witness (MW) must remain identical, as no prior information about the tested state is assumed. In the six-site toy model, there are nine crossing pairs, leaving nine undetermined parameters in $\{a_{ij}\}$. These parameters can be further expressed in terms of $\{m, a_{52}, \cos\theta\}$, a representation that facilitates the calculations detailed in Section~\ref{sec:calculation}.

The classical data set of the 9 pairs from the previous step is:
\begin{equation}
\begin{aligned}
    \left\{d_{ij}\right\}=&
\begin{cases}
    \left\{\langle i\gamma_{1}\gamma_{2}\rangle,\langle i\gamma_{1}\gamma_{4}\rangle,...\right\}, \text{MZM}\\
    \left\{T\langle b_{12}^\dagger b_{12}\rangle,T\langle b_{14}^\dagger b_{14}\rangle,...\right\}, \text{ABS}
\end{cases}\\
    & \text{with } ij\in \{12,14,16,\dots,56\}.
\end{aligned}
\end{equation}
Here $\rho$ represent the projected state copies $\rho_m$ as the input of the parity measurements.
And $T_{ij}$ is a constant related to the tunneling strength between QDs and the sites.
Note that the data come from a set of projected states $\{\rho\}_m$.
From the perturbation calculation in Section~\ref{sec:tunn}, $T_{ij}=(|t_i|^2+|t_j|^2)/t_1^*t_2-t_1t_2^*$.
Further analysis (detailed in Section~\ref{sec:calculation}) reveals that, for MZM states within the odd-parity manifold (a result that also applies to the even-parity manifold), only five of the nine expectation values are nonzero. This outcome follows directly from the conservation of total parity. Consequently, we focus on the five relevant pairs: $\{14,16,34,36,52\}$.

The parity data from $\rho$ are linearly combined with the witness parameters $\{a_{12}, a_{14}, \dots\}$ to compute the trace value:
\begin{equation}
R=\sum_{\left\{ij\right\}=\left\{12,14,...\right\}}a_{ij}d_{ij},
\end{equation}
Theoretical analysis in Section~\ref{sec:witness} predicts the following: if $\rho$ represents an ABS state, the computed value $R$ will be strictly nonnegative. However, if $\rho$ corresponds to an MZM state, there exists a nonzero probability of obtaining a negative value for $R$.

\subsection{Characterizing the Localized Nature of Majorana Zero Modes}

For a set of projected states $\{\rho\}_m$, the success rate of detecting a MZM state, defined as the probability of obtaining a negative witness value $R$, may be relatively low. This limitation arises because the witness value $R$ is not universally negative across the entire MZM subspace. If a projected state does not lie within the negative subspace, the witness fails to identify the MZM state. As detailed in Section~\ref{sec:calculation}, approximately $18\%$ of the MZM subspace yields negative $R$ values under this witness. Consequently, to confidently identify an MZM state, it is often necessary to repeat the protocol across multiple sets of projected states.

To enhance the success rate of MZM detection, the primary steps are repeated across several sets of projected states $\{\rho\}_m$. For an ABS state $\rho$, the witness value $R_m$ is guaranteed to remain positive for all $m$ copies. In contrast, for an MZM state $\rho$, the protocol involves repeating the projection and parity measurement steps until a negative $R_i$ is observed for a specific projected state $\rho_i$. Upon obtaining a negative $R_i$, the state $\rho$ can be identified as an MZM, and $\rho_m$ is retained as the result. If all $m$ sets of projected states yield $R_m \geq 0$, the nature of $\rho$ remains uncertain, necessitating a repetition of the protocol.
Throughout this process, the witness parameters $\{a_{12}, a_{14}, \dots\}$ (or equivalently $\{m, a_{52}, \cos\theta\}$) are kept fixed. Only the parity measurement data are updated for each new set of projected states.



\section{\label{sec:calculation}Calculation on witness performance}

In this section, we give a theoretical calculation of the detection rate of our protocol in a single run.
Here we only consider the MZM states with odd total parity.
For the even parity phase space, the result would be the same.

For the case that an MZM witness Eq.~(\ref{eq:MZMWitness}) acting on an MZM state with odd parity Eq.~(\ref{mzmOdd}), we have the observed value expressed as a function of $\{a_{ij}\}$ and
the MZM state parameters:
\begin{equation}\label{trOdd}
\begin{aligned}
    \operatorname{Tr}(W \rho_{odd})=&1-2 a_{34}(CD-AB)\\
&-a_{36}(-A^{2}-C^{2}+B^{2}+D^{2}) \\
&-2 a_{14}(-AC -BD )-2 a_{16}(BC-AD)\\
&-2 a_{52}(BD-AC)
\end{aligned}
\end{equation}
with all the primes omitted.
And for an MZM state with even parity:
\begin{equation}\label{mzmEven}
    \begin{aligned}
        |\Phi_{even} \rangle =& A'|1_{15}1_{36}0_{24}\rangle + B' |1_{15}0_{36}1_{24}\rangle\\ &+C' |0_{15}1_{36}1_{24}\rangle +D' |0_{15}0_{36}0_{24}\rangle
    \end{aligned}
\end{equation}
Note that ever term in this even-parity expression only differs from the odd one Eq.~(\ref{mzmOdd}) by the first pair.
For this even parity state, the observed value is:
\begin{equation}\label{trEven}
    \begin{aligned}
        \operatorname{Tr}(W \rho_{even})=&1-2 a_{34}(CD-AB)\\
    &+a_{36}(-A^{2}-C^{2}+B^{2}+D^{2}) \\
    &+2 a_{14}(-AC -BD )+2 a_{16}(BC-AD)\\
    &-2 a_{52}(BD-AC)
    \end{aligned}
    \end{equation}

From this we see that with the same parameters $\{a_{ij}\}$, the volume of the
negative subspace is different for odd and even parity space.
So the volume of the negative subspace for different parity is calculated separately.

First, we consider the odd parity space as shown in Eq.~(\ref{mzmOdd}) and Eq.~(\ref{trOdd}).
To calculate the percentage of the negative phase space, we parameterize:
\begin{equation}
\begin{array}{cc}
    a_{14}=&m \cos \theta,\\
    a_{16}=&m \sin \theta,\\
    a_{34}=&-m \sin \theta,\\
    a_{36}=&m \cos \theta.
\end{array}
\end{equation}
and introduce $x = A +C$ and $y = B-D$.

With the new parameters,
\begin{equation}
\begin{aligned}
    \operatorname{Tr}(W\rho_{odd}) =& 1-m \sin \theta \cdot 2xy +m\cos \theta (x^2-y^2) \\&- a_{52}(1-x^2-y^2)
\end{aligned}
\end{equation}
After some algebra, we find the upper and lower bound:
\begin{widetext}
\begin{equation}\label{bound1}
    1+m(x^2 +y^2)-a_{52}(1-x^2-y^2) \geq \operatorname{Tr}(W\rho_{MZM}) \geq
    1-m(x^2 +y^2)-a_{52}(1-x^2-y^2)
\end{equation}
\end{widetext}
From Eq.~(\ref{bound1}) we can see for states which satisfy $x^2+y^2<1$, their
minimal values of $\operatorname{Tr}(W\rho_{MZM})$ are always nonnegative.

We can further parameterize the coefficients of $x$, $y$ terms as:
\begin{equation}
    \operatorname{Tr}(W\rho_{MZM}) = ax^2-by^2-cxy+d
\end{equation}
with $a = m\cos \theta +a_{52}$, $b=m\cos \theta -a_{52}$, 
$c = 2m \sin \theta$, $d = 1-a_{52}$. 
And using the normalized condition of the MZM state, the parameters 
$A, B, C, D$ can also be expressed in terms of trigonometric functions:
\begin{equation}
    \begin{aligned}
        A = &\sin \beta \sin \psi,\\
        B = &\cos \beta \sin \alpha,\\
        C = &\sin \beta \cos \psi,\\
        D = &\cos \beta \cos \alpha.
    \end{aligned}
\end{equation}

By the assumption $a<0$, we can express the inequity 
$\operatorname{Tr}(W\rho_{MZM})<0$ in all the MZM state parameters 
$\{\alpha, \beta, \psi\}$ and witness parameters $\{a, b, c, d\}$:
\begin{widetext}
    \begin{subequations}
        \begin{equation}\label{bound}
            \cos^2 (\beta - \lambda) > \frac{(4ab+c^2)\cos^2 \beta \cos^2(\alpha+\frac{\pi}{4})-2ad}
            {c^2\cos^2 (\alpha + \frac{\pi}{4})+4a^2\cos^2 (\psi-\frac{\pi}{4})},
        \end{equation}
        \begin{equation}\label{lambda}
            \text{with \quad}\cos \lambda = \frac{\sqrt{2}c\cdot \cos(\alpha+\frac{\pi}{4})}
            {[2c^2\cos^2(\alpha+\frac{\pi}{4})+8a^2 \cos^2 (\psi-\frac{\pi}{4})]^{1/2}}.
        \end{equation}  
    \end{subequations}
\end{widetext}
The phase space of all the negative value of $\operatorname{Tr}(W\rho_{MZM})$
is described by Eq.~(\ref{bound}).

To get a numerical result of the detection rate, we set $m=1$, $a_{52}=-1$ and $\cos \theta = -1$.
Then $a=-2, b=0, c = 0, d=2$.
Now we examine weather this set of parameter $\{a_{ij}\}$ satisfies the witness condition. 
The parameter $\{a_{ij}\}$ is:
\begin{equation}
\begin{aligned}
    \mathbf{a} &= \left(\begin{matrix}
            0 & a_{14} & a_{16} \\
            0 & a_{34} & a_{36} \\
            a_{52} & 0 & 0
        \end{matrix}\right) =
        \left(\begin{matrix}
            0 & m cos \theta & msin\theta \\
            0 & -msin\theta & mcos\theta \\
            a_{52} & 0 & 0
        \end{matrix}\right)
        \\&=
    \left(\begin{matrix}
            0 & -1 & 0 \\
            0 & 0 & -1 \\
            -1 & 0 & 0
        \end{matrix}\right)
\end{aligned}
\end{equation}
It is clear that this parameter satisfies both witness conditions for fermions and Andreev bound states.

The bound Eq.~(\ref{bound}) becomes:
\begin{equation}\label{simpleBound}
    \sin^2\beta > \frac{1}{2\cos^2 (\psi -\pi /4)}
\end{equation}
This inequity requires $\cos^2 (\psi-\pi/4)>1/2$, that is
$0<\psi <\pi/2, \pi<\psi<3\pi/2$.

The Jacobian determinant of the 4-dimensional sphere is $\sin \beta \cos \beta$.
So the total phase space volume of MZM states is 
\begin{equation}
    \int_0^{2\pi} d \alpha \int_0^{2\pi} d\psi \int_0^{\pi/2} d\beta \sin \beta \cos \beta = 2\pi^2
\end{equation}

The phase space bounded by Eq.~(\ref{simpleBound}) is
\begin{equation}
    \begin{aligned}
        \int_0^{2\pi} d \alpha \int_0^{\pi/2} \int_\pi^{3\pi/2} d\psi \frac{1}{2}(1-\frac{1}{2\cos^2(\psi-\pi/4)})\\
        =\pi^2-\pi/2
    \end{aligned}
\end{equation}
So the ratio of the negative subspace over the total phase space is
\begin{equation}
    r = \frac{\pi^2-2\pi}{2\pi^2} = 0.1817
\end{equation}

For even parity state we can do a similar calculation, 
and theoretically the negative subspace percentage is the same as the odd subspace, given the same witness parameters $m=1$, $a_{52}=1$, and $\cos \theta = -1$.

\begin{figure*}[htbp]
    \centering
    \subfigure(a)
    {
        \begin{minipage}{8cm}
            \centering
            \includegraphics[width=1\linewidth]{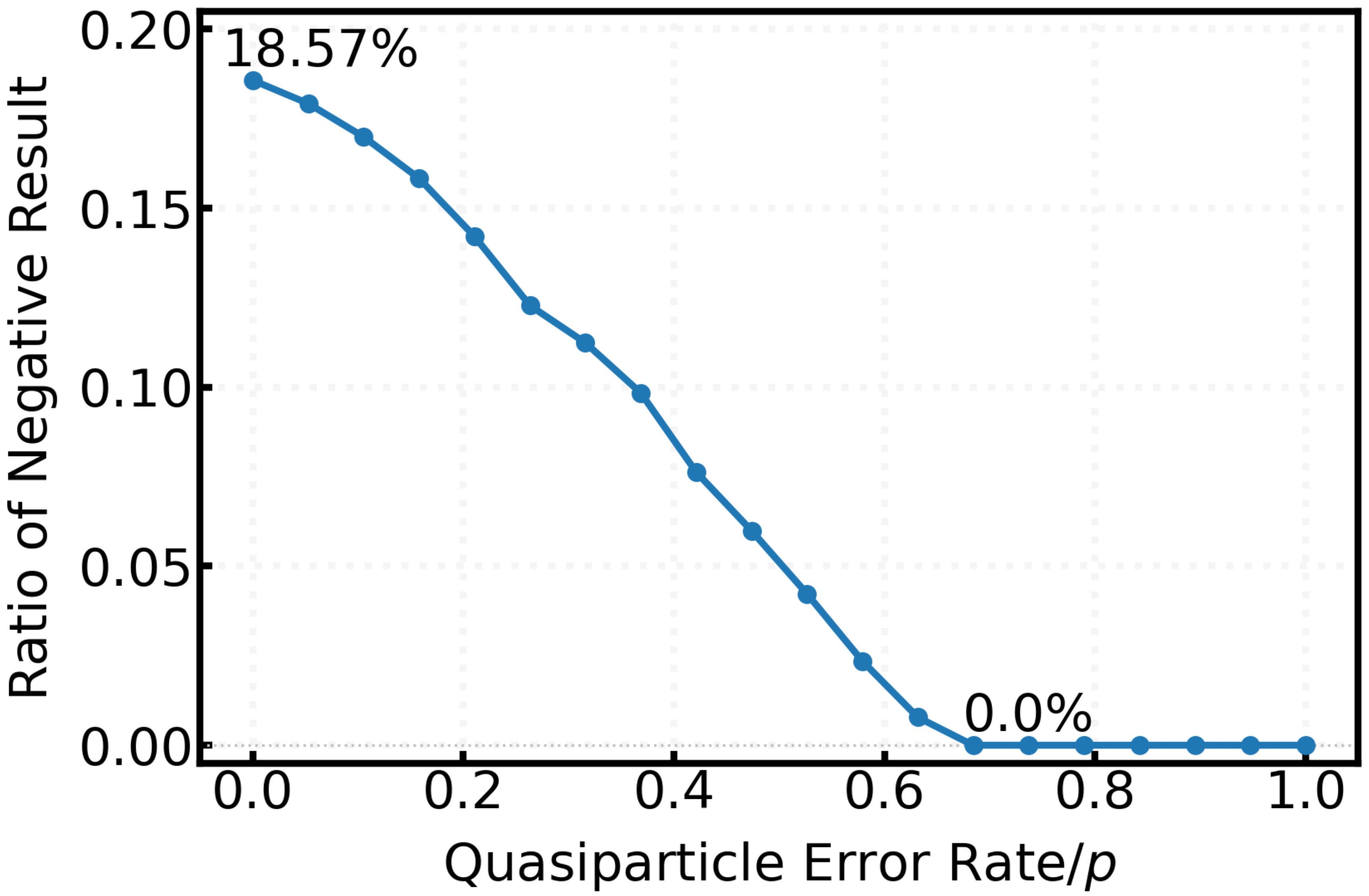}
        \end{minipage}
    }
    \subfigure(b)
    {
        \begin{minipage}{8cm}
            \centering
            \includegraphics[width=1\linewidth]{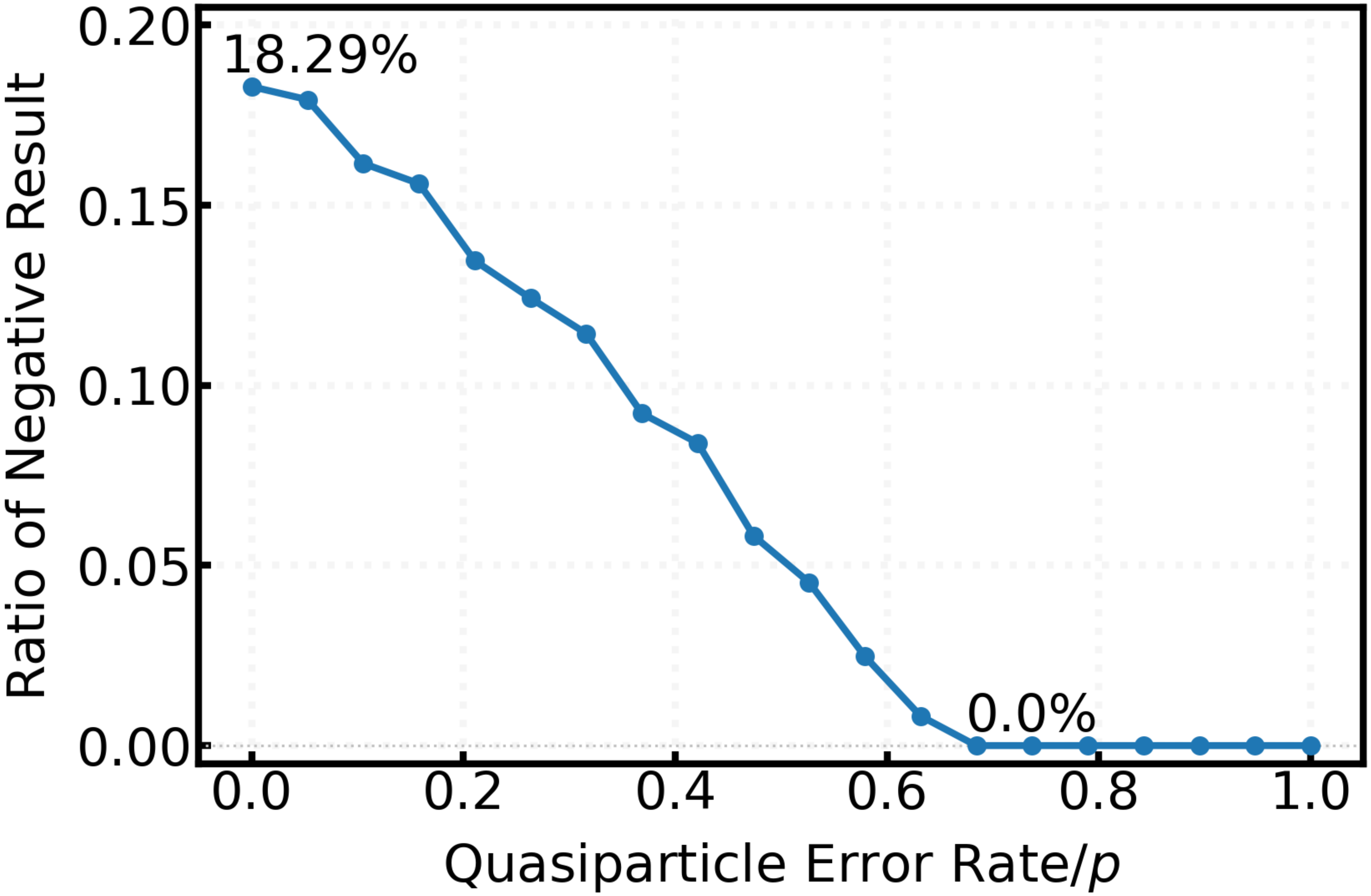}
        \end{minipage}
    }
    \caption{\label{fig:poisoning} The performance of the witness with quasiparticle poisoning. (a) is the result for 10000 random odd-parity MZM states, with witness parameter $\{m=1, a_{52}=-1,\cos\theta=-1\}$; (b) is the counterpart of even-parity MZM states, with witness parameter $a_{52}$ changed into $1$. 
    When $p=0$, there is no quasiparticle poisoning, the ratio of negative space is around the theoretical value; 
    with the increasing of $p$, the ratio of the negative space decreases very fast for both even parity and odd parity cases; 
    at $p=0.6$, the ratio approaches 0, which means the witness no longer works.}
\end{figure*}

\section{\label{sec:MZMerror}Noisy System}
Due to real-world experimental limitations, the preparation of MZM states can not be perfect. Whether the protocol still works for noisy MZM states 
needs more discussion. Since the base of the protocol is the nonlocality
of topological states, local noise that could not affect the topological pairing has little influence on the witness. 

\subsection{The MZM error model}
In this section, we mainly consider the numerical results for quasiparticle
poisoning error in the MZM states.
Yet by the setup in Section~\ref{sec:tunn}, the quasiparticle poisoning 
is suppressed by big $E_C$.
But the longer the wire, the larger the capacity and hence the smaller $E_C$.
So if we want a big $L_W$ to protect the ground state degeneracy,
we must consider quasiparticle events as a type of error in our system.

The quasiparticle poisoning events could take an MZM state from the original parity space to the complementary parity space.
Further, in reality, there are various kinds of errors that
could exist in the experiment.
For example, the overlapping due to the finite size of 
the experimental setup could also introduce imperfection
in the witness process. 
If there is an overlapping between 
two subsystems, then the projection before the parity measurement would leave residual entanglement for ABS state.
After the projection, the witness would turn out to be negative for an ABS state. This would cause the fail of our protocol.

In this paper, we only take the quasiparticle poisoning moise into consideration. And for a more general case of the ABS and MZM hybrid state, see Appendix~\ref{app:hybrid}
.
\subsection{Numerical results of witness performance}
Now numerically we consider the witness performance with the existence of 
quasiparticle (QP) events appear with same possibility $p$ on all sites.
This way all the parity measurements would be affected.
However, how would QP error affect the witness value?
For a fixed set of witness parameters $\{m,\cos\theta,a_{52}\}$, 
the detection rate differs between two complementary parity spaces.
From the previous calculation, with $\{m=1,\cos \theta=-1, a_{52}=-1\}$,
in odd parity space the negative rate is around $18\%$.
Thus if we take a random MZM state in odd parity,
the success rate of this witness is around $18\%$.
However, applying the same witness on MZM states from the even-parity space,
the detection rate is $0\%$, which means this witness is only workable for 
MZM states with odd parity. On the other hand, if the witness is defined by
$\{m=1,\cos \theta=-1, a_{52}=1\}$, then for even parity states, the rate is 
around $18\%$ while for odd parity, the rate is $0\%$.
So by controlling $a_{52}$, the witness could be adjusted to either the odd-parity space or the even-parity space.
Actually, after testing several sets of parameters numerically,
we found that all of the tested parameters are valid only for one of the parity subspaces, but not for the complementary one.
Thus the quasiparticle error would introduce imperfection to the detection, and affect the detection rate of the witness.

Assume the state is in a even parity MZM state.
After the projection, the state $\rho_m$ is poisoned by a quasiparticle channel with the strength $p$:
\begin{equation}
    \mathcal{E}(\rho) = (1-p)\rho + \frac{p}{6}\sum_{i=1}^6 \gamma_i \rho \gamma_i
\end{equation}

Choose 10000 random even-parity MZM states and the witness parameter used in the previous section for for numerical test, the result is as shown in Figure~\ref{fig:poisoning}. From the graphs we can see that for both parity MZM states, the detection rate of the witness goes down as the increasing of the strength of quasiparticle poisoning.

\section{\label{sec:conclusion}Conclusion and outlook}

In topological quantum computing systems, distinguishing nonlocal MZMs from local ABSs remains a significant challenge. This study introduces a protocol designed to detect the nonlocal properties of MZM systems and differentiate MZMs from ABSs by utilizing entanglement witnesses. The proposed approach capitalizes on the nonlocal quantum information intrinsic to MZM pairs, rather than relying solely on direct physical properties. Specifically, within a bipartite system, the protocol requires parity measurements across paired sites that span two subsystems. The resulting measurement data is analyzed using classical computation, providing a necessary condition for determining whether the quantum state exhibits MZM pairing.

In this paper, we present a six-site system containing three MZM pairs as an illustrative example to test our proposed protocol. Our calculations indicate that, under ideal conditions, the overall detection rate for MZM states is approximately $18\%$. Although the theoretical framework suggests that the parity of MZM states does not influence the detection rate, numerical simulations reveal that random quasiparticle poisoning noise significantly impacts the detection rate. 
Moreover, the protocol is effective in more general scenarios, including hybrid states composed of both MZMs and ABSs. In conclusion, this protocol represents a powerful tool for validating MZM pairs in next-generation topological quantum computing devices based on nanowires.

Future directions based on our work include verifying the feasibility of this protocol in experimental systems. Additionally, exploring more efficient quantum information quantities and their implementation schemes in Majorana platforms is crucial for advancing the experimental study of the nonlocal nature of MZMs. While our work demonstrates the potential of such a protocol, practical implementations in real experiments require careful consideration of parameters such as magnetic field strength, tunneling strength, and gate voltages. Furthermore, the impact of noise and measurement inaccuracies in experimental systems must be rigorously addressed. Finally, investigating the intrinsic nature of correlations within MZM pairings presents an intriguing avenue for further research, offering deeper insights into the fundamental properties of these exotic states.

\acknowledgements
This work is supported by the Innovation Program for Quantum Science and Technology (Grant No.~2021ZD0302400), the National Natural Science Foundation of China (Grants No.~92365111), and National Natural Science Foundation of China (Grant
No.~12374158).
The authors would like to thank Ke Ding and Shumeng Chen for helpful discussions.

\appendix
\section{\label{app:EW}Entanglement Witness}

\textbf{Notations and preliminaries}: we denote by $\mathfrak{L}(\mathcal{H})$ the vector space of linear operators acting on a finite dimensional Hilbert space $\mathcal{H}$.\\

For the set of all the positive operators on $\mathfrak{L}(\mathcal{H})$, we denote $\mathfrak{L}_+(\mathcal{H}_{AB})$.\\

The distance on $\mathfrak{L}(\mathcal{H})$ is the Hilbert-Schmidt product:
\begin{equation}
    \langle A| B\rangle_{HS} = \operatorname{Tr}(A^\dagger B)
\end{equation}

The Schmidt decomposition for a state $|\psi\rangle$ in $\mathcal{H}_{AB} = \mathcal{H}_A \otimes \mathcal{H}_B$ is:
\begin{equation}
    |\psi \rangle = \sum_{k=1}^r \mu_k e_k \otimes f_k
\end{equation}
The number of nonzero Schmidt coefficients $\mu_k$ is $r$, and $r$ is called the Schmidt rank. We denote the Schmidt rank $\operatorname{SR}(|\psi\rangle)$. It is an integer between 1 and the dimension $d:= \min{d_A,d_B}$. A state $|\psi\rangle$ is separable iff $\operatorname{SR}(|\psi\rangle) = 1$. And if $\operatorname{SR}(|\psi\rangle) = d$, $|\psi\rangle$ is called maximally entangled.

The notations are taken the same as in~\cite{chruscinski_entanglement_2014}.

\subsection{The definition of an EW}
In a review paper~\cite{chruscinski_entanglement_2014}, the witness is defined as an operator $W \in \mathfrak{L}(H_{AB})$  is an EW if and only if it is \textit{block-positive} but not positive.

\textbf{Block-positive}:
A Hermitian operator $W\in \mathfrak{L}(H_{AB})$ is block-positive if for all product states in $\mathfrak{L}(H_{AB})$, $W$ is positive:
\begin{equation}
    \langle \psi \otimes \phi |W| \psi \otimes \phi \rangle \geq 0
\end{equation}
The set of block-positive operators are a convex cone in $\mathfrak{L}(\mathcal{H}_{AB})$, and is denoted $\mathbb{W}_1$. The set of EW is then $\mathbb{W}_1 - \mathfrak{L}(\mathcal{H}_{AB})$.

The subscript in $\mathbb{W}_1$ correspond to the Schmidt rank of the product state, which is $1$. Further, we can also define k-block-positive sets using $\operatorname{SR}$:

\textbf{k-Block-positive}: the set of Hermitian operators that are positive over the set of states with SR at most $k$:
\begin{equation}
    \mathbb{W}_k=\left\{W \in \mathfrak{L}\left(\mathcal{H}_{A B}\right) \mid\langle\Psi|W| \Psi\rangle \geqslant 0 ; \quad \operatorname{SR}(\Psi) \leqslant k\right\}
\end{equation}

These convex cones have a relation of chain inclusion:
\begin{equation}
    \mathfrak{L}_{+}\left(\mathcal{H}_{A B}\right)=\mathbb{W}_d \subset \mathbb{W}_{d-1} \subset \cdots \subset \mathbb{W}_1 .
\end{equation}

Then any EW can be classified depending on which subset $W$ sets in the chain: $W$ is a $k$-Schmidt witness $W \in \mathbb{W}_{k-1} - \mathbb{W}_k$.

\subsection{Banach separation theorem: The completeness of witnesses}
For any entangled states, there exist an EW separate it fram the convex set of product states. More precisely, we have the Lemma:

\textbf{Lemma 1\cite{horodecki_separability_1996}}: For any entangled state $\rho \in \mathcal{H}_{AB}$, there exists a Hermitian $W$ such that
\begin{equation}
    \operatorname{Tr} (W\rho) \quad \text{and} \quad \operatorname{Tr} (W\sigma) \quad \text{for all separable $\sigma$.}
\end{equation}

\textit{Proof}: This lemma can be proved by the Hahn-Banach theorem. The theorem can be stated as: 

\textbf{Hahn-Banach Theorem}: for two convex closed sets $W_1,W_2$ in a real Banach space, if one of them is also compact, then there exist a continuous functional $f$ and a real $\alpha$ such that for any $w_1\in W_1$ and $w_2 \in W_2$, $f(w_1)< \alpha \leq f(w_2)$.

The Hilbert space any dimension is a Banach space. And with the fact that a set with one element is compact, we can get the above Lemma by defining the continuous functional to be $f(\rho) = \operatorname{Tr}(W\rho)$.

\subsection{Optimal EW and the formation of EW}
The form of EW used in the paper comes from ~\cite{Yu05}.
However the discussion in~\cite{Yu05} is not very clear.
Here the optimization conditions for any EW, 
as well as a closely related form of EW, which is based on realignment criterion, would be discussed first.
This part comes from a highly-cited review paper~\cite{chruscinski_entanglement_2014}.

\subsubsection{Optimal EWs}

For an EW $W$, the set of all the entangled states that can be detected by W is:
\begin{equation}
    D_W = \{\rho \in \mathfrak{G}(\mathcal{H}_{AB})|\operatorname{Tr}(\rho W) <0 \}
\end{equation}
Here $\mathfrak{G}(\mathcal{H}_{AB})$ is the mixed state space on the Hilbert space of the system $AB$.

With this we can say an EW $W_1$ is finer than $W_2$, if $D_{W_1} \supseteq D_{W_2}$. An EW is called optimal on $\mathfrak{G}(\mathcal{H}_{AB})$, if there is no other EW finer than $W$.

Note that the set of block-positive operators is included in the set of positive operators. And if $W_1$ is finer than $W_2$, then there exist a positive $P$ s.t. $W_2 = W_1 +P$. Then we can see $W$ is the optimal EW iff for any positive operator $P$, $W-P$ is no longer block-positive. 

A way to construct an EW from the positive operators is:
\begin{equation}
    W_\lambda = P - \lambda \mathbb{I}_A \otimes \mathbb{I}_B
\end{equation}
Here $P\in \mathfrak{L}_+(\mathcal{H}_{AB})$ is a positive operator. $W$ is block-positive if $\lambda \leq \lambda_0 = \inf \langle \alpha \otimes \beta | P | \alpha \otimes \beta\rangle$. With $\lambda = \lambda_0$, the witness $W_{\lambda_0}$ is not necessarily optimal according to the previous definition. However, it is located on the boundary of the set of block-positive. It is called \textit{weakly optimal} in some contexts.
In other words, if we find a separable state $\sigma$ s.t. $\operatorname{Tr} (W\sigma) = 0$, then the witness $W$ is weak optimal. 

For optimal EWs, ~\cite{lewenstein_optimization_2000} gives the proposition:

\textbf{Proposition}: If we can find a set of orthogonal separable states $\{\sigma\}$ that satisfy $\operatorname{Tr} (W\sigma) = 0$ and they \textit{span the Hilbert space}, then the witness is optimal.

\subsubsection{The EW from CCNR criterion}\label{sec:CCNR}
For the Schmidt decomposition of a density matrix $\rho$, the following computable cross norm or realignment (CCNR) criterion of holds:

\textbf{Theorem} (CCNR criterion):\cite{guhne_entanglement_2009}
The state $\rho$ has the Schmidt decomposition:
\begin{equation}
    \rho = \sum_k \lambda_k G_k^A \otimes G_k^B
\end{equation}
The Schmidt coefficients $\{\lambda_k\}$ can give the separability:
if $\rho$ is separable then $\sum_k \lambda_k \leq 1$. In other words, $\sum_k \lambda_k >1$ is the necessary condition for $\rho$ to be entangled.

Inspired by the CCNR criterion, we can construct a EW:
\begin{equation}\label{eq:CCNR}
    W = \mathbb{I}_A \otimes \mathbb{I}_B -\sum_k G_k^A \otimes G_k^B
\end{equation}
We can proof the block-positivity. For product states,
\begin{equation}
    \langle \alpha \otimes \beta | W | \alpha \otimes \beta\rangle = 1 - \sum_k \langle \alpha | G_k^A | \alpha \rangle \langle \beta | G_k^B | \beta \rangle
\end{equation}
Since $G_k^{A/B}$ is complete basis of the corresponding subsystem, we have:
\begin{equation}
    |\alpha\rangle \langle\alpha |=\sum_k a_k G_k^A, \quad | \beta \rangle\langle\beta|=\sum_l b_l G_l^B
\end{equation}
And with the normalization condition, $\sum_k a_k^2 = \sum_l b_l^2 = 1$. With Cauchy-Schwarz inequality, $\sum_k a_k b_k \leq 1$. Then we get:
\begin{equation}
    \langle \alpha \otimes \beta | W | \alpha \otimes \beta\rangle = 1 - \sum_k a_k b_k \geq 0.
\end{equation}
For an entangled state that can be detected by this EW, the trace would be negative:
\begin{equation}
    \operatorname{Tr}(W\rho) = 1-\sum_k \lambda_k <0
\end{equation}
So $W$ defined in Eq.~(\ref{eq:CCNR}) is an EW.

\section{\label{app:Perturbation}The Perturbation Calculation of Tunneling}
As in Eq.~(\ref{eq:totalH}), the Hamiltonian of the system is
\begin{equation}
\begin{aligned}
    H & = H_0 + H_{QD} + H_{tunn}\\
    & =  H_{\text{BCS}} + E_C (N_s - N_g)^2 + h\hat{n}_f + \epsilon_c(\hat{n}_f - n_g)^2 +H_{tunn}.
\end{aligned}
\end{equation}
The form of the tunneling term depends on the actual state on the sites of the superconducting island.

\subsection{\label{app:MZMPerturbation} The Majorana zero mode tunneling}
For the Majorana zero mode state, the tunneling term in the Hamiltonian is the same as Eq.~(\ref{eq:MZMtunn})
\begin{equation}
    H_{tunn,MZM}^{(1,2)} = -\mathrm{i}\frac{e^{-\mathrm{i}\phi/2}}{2}\left( t_1 f_1^{\dagger} \gamma_1+ t_2 f_1^{\dagger} \gamma_2 \right)+\text { H.c. }
\end{equation}
According to the perturbation theory, the second order perturbation of the ground state energy is
\begin{equation}
    E_g = E_g^0 + V_{gg} +\frac{V_{ge}V_{eg}}{E_g^0 - E_e^0}.
\end{equation}
The subscript $g$ represents the ground state and the subscript $e$ represents the excited state.
Consider the condition that the QD is occupied at first and the tunneling is from the QD to the island. The ground state is $|N_S = 0\rangle \otimes |n_f = 1\rangle$. The excited state is $|N_S = 1\rangle \otimes |n_f = 0\rangle$.
The matrix element $V_{eg}$ is
\begin{equation}
    \begin{aligned}
        V_{eg} & = \langle e | V | g \rangle\\
        & = \langle N_S = 1 |\otimes\langle n_f = 0| -\mathrm{i}\frac{e^{-\mathrm{i}\phi/2}}{2}\left( t_1 f_1^{\dagger} \gamma_1+ t_2 f_1^{\dagger} \gamma_2 \right)\\
        & \quad + \mathrm{i}\frac{e^{\mathrm{i}\phi/2}}{2}\left( t_1^* \gamma_1 f_1 + t_2^* \gamma_2 f_1  \right)|N_S = 0\rangle |n_f = 1\rangle\\
        & = \frac{\mathrm{i}}{2} \langle N_S = 1 |\otimes\langle n_f = 0|e^{\mathrm{i}\phi/2}(t_1^* \gamma_1f + t_2^* \gamma_2f) \\
        & \quad |N_S = 0\rangle |n_f = 1\rangle\\
        & = -\frac{\mathrm{i}}{2} \langle N_S = 1 |\otimes\langle n_f = 0|f(t_1^* \gamma_1 + t_2^* \gamma_2) \\
        & \quad |N_S = 1\rangle |n_f = 1\rangle\\
        & = -\frac{\mathrm{i}}{2} (t_1^* \gamma_1 +t_2^* \gamma_2).
    \end{aligned}
\end{equation}

Writing the QD energy of occupancy 0 and 1 as $\epsilon_0 = \epsilon_C n_g^2$ and $\epsilon_1 = \epsilon_C (1-n_g^2) +h$, the total energy under the perturbation is
\begin{equation}
\begin{aligned}
     \varepsilon_1^{\mathrm{MZM}}& =E_C N_g^2+\epsilon_1 + \frac{V_{ge}V_{eg}}{E_g^0 - E_e^0}\\
     & = E_C N_g^2+\epsilon_1 - \frac{(t_1\gamma_1+t_2\gamma_2)(t_1^*\gamma_1+t_2^*\gamma_2)}{4\left[E_C\left(1-2 N_g\right)+\epsilon_0-\epsilon_1\right]}\\
     & = E_C N_g^2+\epsilon_1-\frac{\left|t_1\right|^2+\left|t_2\right|^2+\mathrm{i} \left(t_1^* t_2-t_1 t_2^*\right)\mathrm{i}\gamma_1\gamma_2}{4\left[E_C\left(1-2 N_g\right)+\epsilon_0-\epsilon_1\right]}\\
     & = E_C N_g^2+\epsilon_1-\frac{\left|t_1\right|^2+\left|t_2\right|^2+\mathrm{i} p_{12}\left(t_1^* t_2-t_1 t_2^*\right)}{4\left[E_C\left(1-2 N_g\right)+\epsilon_0-\epsilon_1\right]}.
\end{aligned}
\end{equation}

Now we consider the initial state as $|N_S = 0\rangle \otimes |n_f = 0\rangle$, and the state after tunneling is $|N_S = -1\rangle \otimes |n_f = 1\rangle$. The total energy with perturbation can be obtained in the same way.

\begin{equation}
    \varepsilon_0^{\mathrm{MZM}} = E_C N_g^2+\epsilon_0-\frac{\left|t_1\right|^2+\left|t_2\right|^2-\mathrm{i} p_{12}\left(t_1^* t_2-t_1 t_2^*\right)}{4\left[E_C\left(1+2 N_g\right)+\epsilon_1-\epsilon_0\right]}.
\end{equation}

\subsection{\label{app:FPerturbation} The fermionic state tunneling}
Now we consider the fermionic case, which is a special case of ABS. The tunneling term is the same as Eq.~(\ref{eq:Ftunn})
\begin{equation}
     H_{tunn,F}^{(1,2)} = -\frac{\mathrm{i}e^{-\mathrm{i}\phi/2}}{2}\left(t_1 f_1^{\dagger} c_1+ t_2 f_1^{\dagger} c_2 \right)+\text { H.c. }
\end{equation}

For the ground state $|N_S = 0\rangle \otimes |n_f = 1\rangle$ and the excited state $|N_S = 1\rangle \otimes |n_f = 0\rangle$, the same calculation of the matrix element in second order perturbation gives
\begin{equation}
\begin{aligned}
    V_{eg} & = \langle e | V | g \rangle \\
    & = \langle N_S = 1|\otimes \langle n_f = 0| -\frac{\mathrm{i}e^{-\mathrm{i}\phi/2}}{2}\left(t_1 f_1^{\dagger} c_1+ t_2 f_1^{\dagger} c_2 \right) \\
    &\quad + \frac{\mathrm{i}e^{\mathrm{i}\phi/2}}{2}\left(t_1^* c_1^\dagger f_1 + t_2^* c_2^{\dagger} f_1  \right)|N_S = 0\rangle \otimes |n_f = 1\rangle\\
    & = \frac{\mathrm{i}}{2} \langle N_S = 1|\otimes \langle n_f = 0|e^{\mathrm{i}\phi/2}(t_1^*c_1^\dagger f_1 + t_2^* c_2^\dagger f_1)\\
    & \quad |N_S = 0\rangle \otimes |n_f = 1\rangle\\
    & = -\frac{\mathrm{i}}{2} \langle N_S = 1|\otimes \langle n_f = 0|f_1(t_1^*c_1^\dagger + t_2^*c_2^\dagger)\\
    & \quad |N_S = 1\rangle \otimes |n_f = 1\rangle\\
    & = -\frac{\mathrm{i}}{2}(t_1^*c_1^\dagger + t_2^*c_2^\dagger)
\end{aligned}
\end{equation}
Note that $V_{ge}V_{eg}$ can be represented by the unitary transformation of fermion operators
\begin{equation}
    \begin{aligned}
        V_{ge}V_{eg} &= \frac{1}{4}(t_1c_1 + t_2c_2)(t_1^*c_1^\dagger + t_2^*c_2^\dagger) \\
        & = \frac{1}{4} (|t_1|^2 c_1c_1^\dagger + |t_2|^2 c_2c_2^\dagger+t_1t_2^*c_1c_2^\dagger+t_2t_1^*c_2c_1^\dagger)\\
        & = \frac{1}{4}[|t_1|^2 (1-c_1^\dagger c_1)+|t_2|^2(1-c_2^\dagger c_2)\\
        & \quad -(t_1 t_2^*c_1c_2^\dagger+t_2t_1^* c_1^\dagger c_2)] \\
        & = \frac{1}{4}(|t_1|^2 + |t_2|^2) - \frac{1}{4} (|t_1|^2 c_1^\dagger c_1+|t_2|^2c_2^\dagger c_2\\
        & \quad +t_1 t_2^*c_1c_2^\dagger+t_2t_1^* c_1^\dagger c_2)\\
        & = \frac{1}{4}(|t_1|^2 + |t_2|^2)(1-b_{12}^\dagger b_{12}).
    \end{aligned}
\end{equation}
Here $b_{12}\equiv b_{12}^0= (t_1 c_1 + t_2 c_2)/\sqrt{\left|t_1\right|^2+\left|t_2\right|^2}
$ is a unitary transformed basis of fermion operator basis
\begin{equation}
    \left(\begin{matrix}
        b_{12}^0\\
        b_{12}^1
    \end{matrix}\right)=\frac{1}{\sqrt{|t_1|^2+|t_2|^2}}
    \left(\begin{matrix}
        t_1 & t_2 \\
        1 & 0
    \end{matrix}\right)
    \left(\begin{matrix}
        c_1\\
        c_2
    \end{matrix}\right).
\end{equation}

So the ground state energy for fermionic state under perturbation is 

\begin{equation}
        \varepsilon_1^F  =E_C N_g^2+\epsilon_1-\frac{\left|t_1\right|^2+\left|t_2\right|^2-\left(\left|t_1\right|^2+\left|t_2\right|^2\right) b_{12}^{\dagger} b_{12}}{4\left[E_C\left(1-2 N_g\right)+\epsilon_0-\epsilon_1\right]} 
\end{equation}

Applying the same calculation to the ground state $|N_S = 0\rangle \otimes |n_f = 0\rangle$ and the excited state $|N_S = -1\rangle \otimes |n_f = 1\rangle$, one has the perturbed energy

\begin{equation}
    \varepsilon_0^F =E_C N_g^2+\epsilon_0-\frac{\left|t_1\right|^2+\left|t_2\right|^2+\left(\left|t_1\right|^2+\left|t_2\right|^2\right) b_{12}^{\dagger} b_{12}}{4\left[E_C\left(1+2 N_g\right)+\epsilon_1-\epsilon_0\right]}
\end{equation}

\subsection{\label{app:APerturbation} The Andreev bound state tunneling}
As for the Andreev bound state, the tunneling term in the total Hamiltonian is
\begin{equation}
\begin{aligned}
     H_{tunn,ABS}^{(1,2)} = &-\frac{\mathrm{i}e^{-\mathrm{i}\phi/2}}{2}[t_1 f_1^{\dagger} (\left| u_1\right| c_1 + \left| v_1\right| c_1^\dagger)\\
    &+ t_2 f_1^{\dagger} (\left| u_2\right| c_2 + \left| v_2\right| c_2^\dagger)]+\text { H.c. }
\end{aligned}
\end{equation}

Note that the only difference between the ABS tunneling Hamiltonian and the fermion tunneling Hamiltonian Eq.~(\ref{eq:Ftunn}) is replacing the fermion creation (or annihilation) operators by the linear combination of fermion creation and annihilation operators.
\begin{equation}\label{eq:app_replacing}
    c_i \to \alpha_i = |u_i|c_i+|v_i|c_i^\dagger
\end{equation}

Instead of using the Fock states $|0\rangle , |1\rangle$, we can use the Fock states of the quasiparticle operator as the perturbation basis.
The anti-commutation relation $\alpha_i^\dagger \alpha_j = -\alpha^\dagger_i \alpha_j$ and $\alpha_i \alpha_i^\dagger = 1-\alpha_i^\dagger \alpha_i$ still applies to each term in the ABS operator.
Therefore, the perturbation calculation is the same and the perturbed ground state energy in the ABS case can be obtained by directly performing the replacement Eq.~(\ref{eq:app_replacing}). And the effective measurement is $b_{12,ABS}^\dagger b_{12,ABS}$, with
\begin{equation}
\begin{aligned}
    b_{12,ABS}^\dagger &= \frac{t_1^* \alpha_1^\dagger + t_2^* \alpha_2^\dagger}{\sqrt{\left|t_1\right|^2+\left|t_2\right|^2}}\\
    & = \frac{t_1^* (\left| u_1\right| c_1^\dagger + \left| v_1\right| c_1) + t_2^* (\left| u_2\right| c_2^\dagger + \left| v_2\right| c_2)}{\sqrt{\left|t_1\right|^2+\left|t_2\right|^2}}.
\end{aligned}
\end{equation}

\section{\label{app:AWCandidate}The Witness Candidate Condition}
The fermionic state is a special case of Andreev bound states.
In this section we first give the witness candidate condition for the fermionic state. Then we move on to the more general Andreev bound state case in a similar way. 

\subsection{The witness candidate condition of fermionic witness}
The result in Section~\ref{sec:tunn} as well as in Appendix~\ref{app:FPerturbation} shows that two site measurement in fermionic state is $b_{12}^\dagger b_{12}$, with

\begin{equation}
    b_{12}^\dagger = \frac{t_1^* c_1^\dagger + t_2^* c_2^\dagger}{\sqrt{\left|t_1\right|^2+\left|t_2\right|^2}}
\end{equation}

The measurement outcome on a pair of sites, for example $1$ and $2$, is
\begin{widetext}
    \begin{equation}
        \begin{aligned}
            \operatorname{Tr} [b_{12}^\dagger b_{12} \rho] &= \sum_x \langle x|b_{12}^\dagger b_{12}(\rho_1 \otimes \rho_2)| x \rangle\\
            &= \frac{1}{\left|t_1\right|^2+\left|t_2\right|^2}\sum_x \langle x|(|t_1|^2c_1^\dagger c_1 + t_1^* t_2 c_1^\dagger c_2 +t_2^* t_1 c_2^\dagger c_1 +|t_2|^2c_2^\dagger c_2)(\rho_1 \otimes \rho_2)| x \rangle
        \end{aligned}
    \end{equation}
\end{widetext}

The trace can be taken separately on two sites, since $\rho = \rho_1 \otimes \rho_2$ is separable. We denote trace on site $1$ by $\langle\cdot\rangle_1$ and trace on site $2$ $\langle\cdot\rangle_2$. Then we get
\begin{widetext}
    \begin{equation}\label{eq:site12}
\begin{aligned}
    \operatorname{Tr} [b_{12}^\dagger b_{12} (\rho_1 \otimes \rho_2)] &= \frac{|t_1|^2 \langle c_1^\dagger c_1\rangle_1 + t_1^* t_2 \langle c_1^\dagger\rangle_1 \langle c_2\rangle_2 +t_2^* t_1 \langle c_2^\dagger\rangle_2 \langle c_1\rangle_1 +|t_2|^2 \langle c _2^\dagger c_2\rangle_2}{\left|t_1\right|^2+\left|t_2\right|^2}\\
    &\leq \frac{|t_1|^2 +|t_2|^2 + t_1^* t_2 \langle c_1^\dagger\rangle_1 \langle c_2\rangle_2 +t_2^* t_1 \langle c_2^\dagger\rangle_2 \langle c_1\rangle_1}{\left|t_1\right|^2+\left|t_2\right|^2}
\end{aligned}
\end{equation}
\end{widetext}

To see the crossing terms, we can change to the Pauli basis.
The transformation between fermionic operators and Pauli operators is
\begin{equation}
\begin{array}{c}
    \begin{array}{cc}
        c_1^\dagger = \frac{1}{2}(\sigma_1^x + i\sigma_1^y), & c_2^\dagger = \frac{(-1)^{c_1^\dagger c_1}}{2}(\sigma_2^x + i\sigma_2^y), \\
        c_1 = \frac{1}{2}(\sigma_1^x - i\sigma_1^y), & c_2 = \frac{(-1)^{c_1^\dagger c_1}}{2}(\sigma_2^x - i\sigma_2^y),\\
    \end{array} \\
    2c_i^\dagger c_i - 1 = \sigma_i^z. 
\end{array}
\end{equation}
And the state can also be expressed in Pauli basis $\rho_i = \sum_{j} p_i \sigma_i^j$, $\sum_j p_{ij}^2 = 0$.
Then the trace on each site is
\begin{widetext}
    \begin{equation}
    \begin{aligned}
        \langle (-1)^{c_1^\dagger c_1}c_1^\dagger \rangle_1 &= \operatorname{Tr}(c_1^\dagger\rho_1)\\
        & = \operatorname{Tr}[(-1)^{c_1^\dagger c_1}c_1^\dagger(p_{10} I + p_{11} \sigma^x + p_{12}\sigma^y +p_{13} \sigma^z)]\\
        & = \operatorname{Tr}[\left(\begin{array}{cc}
            -1 & 0 \\
             0& 1
        \end{array}\right)\cdot \left(\begin{array}{cc}
            0 & 1 \\
             0& 0
        \end{array}\right)(p_{10} I + p_{11} \sigma^x + p_{12}\sigma^y +p_{13} \sigma^z)]\\
        & = -(p_{11} + ip_{12})
    \end{aligned}
\end{equation}
\begin{equation}
    \begin{aligned}
        \langle (-1)^{c_1^\dagger c_1}c_1 \rangle_1 &= p_{11} - ip_{12} = -\langle (-1)^{c_1^\dagger c_1} c_1^\dagger \rangle_1^*,\\
        \langle c_2^\dagger \rangle_2 &= p_{21} + ip_{22},\\
        \langle c_2 \rangle_2 &= p_{21} - ip_{22} = \langle c_2^\dagger \rangle_2^*
    \end{aligned}
\end{equation}

\end{widetext}

So the crossing terms on site $1$ and site $2$ is real

\begin{widetext}
    \begin{equation}
    \begin{aligned}
        t_1^* t_2 \langle c_1^\dagger\rangle_1 \langle c_2\rangle_2 +t_2^* t_1  \langle c_2^\dagger\rangle_2 \langle c_1\rangle_1
        &= -t_1^* t_2 \langle c_2\rangle_2 \langle c_1^\dagger\rangle_1 +t_2^* t_1  \langle c_2^\dagger\rangle_2 \langle c_1\rangle_1\\
        &= t_1^* t_2 (p_{11} + ip_{12})(p_{21} - ip_{22}) + \text{conj.})\\
        &= 2[\operatorname{Re}(t_1^* t_2)(p_{11}p_{21} + p_{12}p_{22}) - \operatorname{Im}(t_1^* t_2)(p_{12}p_{21}-p_{11}p_{22})]\\
        &\leq 2\sqrt{\operatorname{Re}(t_1^* t_2)^2+\operatorname{Im}(t_1^* t_2)^2}=2|t_1^* t_2|^2
    \end{aligned}
\end{equation}
\end{widetext}

The last inequality takes equality if $p_{ij}$ are all equal to $1/2$.
So we bound the parameter $a$ with the tunneling $t_1$, $t_2$ as:
\begin{equation}
    \operatorname{Tr} [b_{12}^\dagger b_{12} (\rho_1 \otimes \rho_2)] \leq a_{12} \frac{|t_1|^2 +|t_2|^2 +  2|t_1^* t_2|}{\left|t_1\right|^2+\left|t_2\right|^2}.
\end{equation}

Then the observed value for $W_F$ on separable state is
\begin{equation}
\begin{aligned}
    \operatorname{Tr} [W_F (\rho_A \otimes \rho_B)] & \geq 1- \sum_{\langle i,j\rangle}a_{ij} \left(1+ \frac{2|t_i^* t_j|}{\left|t_i\right|^2+\left|t_j\right|^2}\right)\\
    & \equiv 1-\sum_{\langle i,j\rangle}a_{ij}\cdot T_{ij}(t_i.t_j).
\end{aligned}
\end{equation}

This inequality is the condition for $\{a_{ij}\}$ in fermion witness $W_F$. For $\forall t_i,t_j \in \mathbb{C}$, the function of tunneling coefficient has upper and lower bounds, $0 \leq T_{ij}(t_i.t_j)\leq 1$. 

\subsection{The witness candidate condition of ABS witness}

The fermion state is a special case of Andreev bound state. For the case of ABS, the tunneling of particle and hole on each site can have different strength.
The operators measured in the witness would be in terms of $b_{ij,ABS}^\dagger b_{ij,ABS}$, which is showed in Eq.~(\ref{eq:abs_b}).
And the condition of EW candidate Eq.~(\ref{eq:EWCondition}) is then calculated to give a bound on $\{a_{ij}\}$.

In the calculation, there would be terms like $\langle c_i^\dagger c_j^\dagger \rangle$ and $\langle c_i c_j \rangle$. The former corresponds to adding a Cooper pair to the island, and the later removing one. So we take the average to be $e^{i\phi}$ and $e^{-i\phi}$, with $\phi$ the overall superconducting phase on the island.

The result for $W_A$ is:
\begin{equation}
\begin{aligned}
    \operatorname{Tr} [W_A (\rho_1 \otimes \rho_2)] &\geq 1- \sum_{\langle i,j\rangle}a_{ij} (1+\sqrt{2}+\frac{2|t_i^* t_j|}{\left|t_i\right|^2+\left|t_j\right|^2})\\
    &\equiv1- \sum_{\langle i,j\rangle}a_{ij}\cdot (\sqrt{2} + T_{ij}(t_i.t_j)).
\end{aligned}
\end{equation}

\section{\label{app:hybrid}More general cases in our protocol}
We have shown detailed probability of successfully distinguishing two system in one-shot experiment in ideal case. And we have ignored the cases where there's mixture of MZM and fermion in one experiment. For this case, we will take a form of density matrix like:
\begin{equation}
\rho=\rho_{MZM}\otimes\rho_{ABS}
\end{equation}
The state on, for example, site 5 and 6 is ABS, while the state on sites 1-4 is MZMs. Then the witness value on such state is:
\begin{equation}
   \begin{aligned}
       Tr(W\rho)&=Tr(\rho)-Tr(W_{1-4}\rho_{MZM})Tr(W_{5-6}*\rho_{fer})\\
       &=1-Tr(W_{1-4}*\rho_{MZM})Tr(W_{5-6}*\rho_{ABS})
   \end{aligned} 
\end{equation}

In a smaller subspace of the original space, the positivity of the witness on fermions isn't changed.
As soon as we detect negativity then we can judge that there are MZMs in the system.

\bibliographystyle{apsrev4-1}
\bibliography{refs}

\end{document}